# Cyberswarm: a novel swarm intelligence algorithm inspired by cyber community dynamics

Abdelsadeq Elfergany[1] · Ammar Adl[1] · Mohammed Kayed[1]

**Abstract** Recommendation systems face challenges in dynamically adapting to evolving user preferences and interactions within complex social networks. Traditional approaches often fail to account for the intricate interactions within cyber-social systems and lack the flexibility to generalize across diverse domains, highlighting the need for more adaptive and versatile solutions. In this work, we introduce a general-purpose swarm intelligence algorithm for recommendation systems, designed to adapt seamlessly to varying applications. It was inspired by social psychology principles. The framework models user preferences and community influences within a dynamic hypergraph structure. It leverages centrality-based feature extraction and Node2Vec embeddings. Preference evolution is guided by message-passing mechanisms and hierarchical graph modeling, enabling real-time adaptation to changing behaviors. Experimental evaluations demonstrated the algorithm's superior performance in various recommendation tasks, including social networks and content discovery. Key metrics such as Hit Rate (HR), Mean Reciprocal Rank (MRR), and Normalized Discounted Cumulative Gain (NDCG) consistently outperformed baseline methods across multiple datasets. The model's adaptability to dynamic environments allowed for contextually relevant and precise recommendations. The proposed algorithm represents an advancement in recommendation systems by bridging individual preferences and community influences. Its general-purpose design enables applications in diverse domains, including social graphs, personalized learning, and medical graphs. This work highlights the potential of integrating swarm intelligence with network dynamics to address complex optimization challenges in recommendation systems.
**Graphic abstract**

Extended author information available on the last page of the article

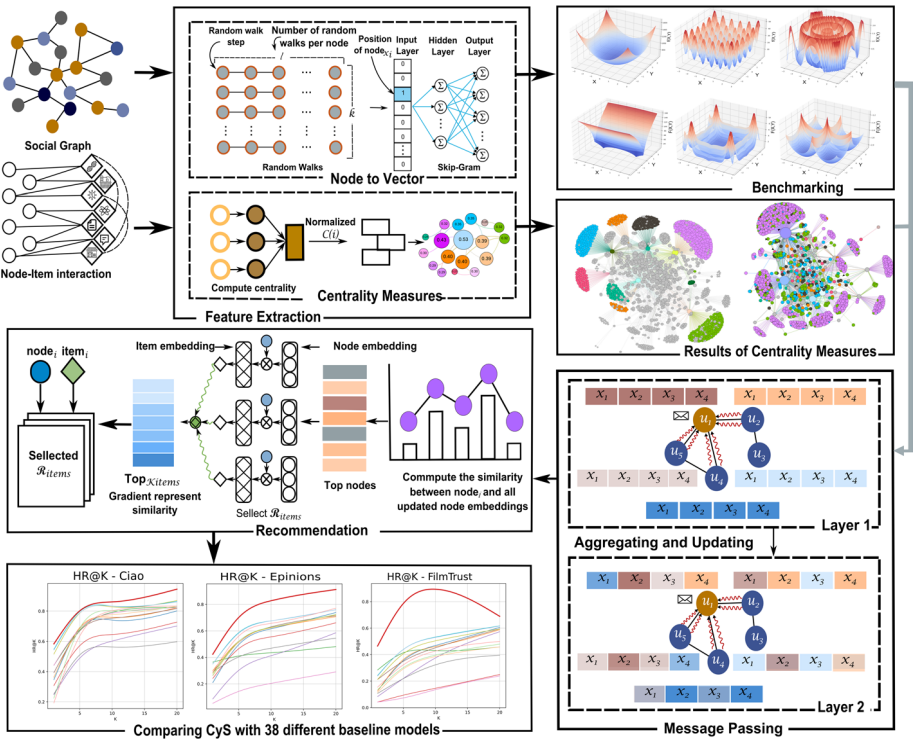

# 1 Introduction

In this paper, we present our CyberSwarm Algorithm (CyS). It is a general-purpose swarm intelligence algorithm inspired by the interactive dynamics of social networks. It models each entity as a node within a graph, where edges represent connections like shared interests or exchanges of information, facilitating dynamic, context-sensitive interactions. The proposed model is guided by principles from social psychology, incorporating aspects of attitude and belief evolution. Drawing on theories such as Social Judgment and Consistency Theory, each node's preferences evolve as a function of its interactions. Starting from an initial preference vector $P_i(0)$, each node iteratively adjusts its preferences toward alignment with its social context. This adaptive process is encapsulated in a time-dependent preference vector $P_i(t+1)$, which integrates influence from neighboring nodes. The result is a flexible recommendation framework that adapts well to various applications.

Our proposed framework incorporates a hypergraph ($\mathcal{H}$) to enhance its performance further. Each node's feature vector is enriched by centrality measures ($\mathcal{M}_{\text{central}}(\mathcal{H})$) and Node2Vec embeddings. These features play a crucial role in capturing evolving preferences in real time. A central aspect of the proposed algorithm is the Dynamic Collaborative Swarm

Equilibrium (DCSE) theorem. The DCSE theorem explains how equilibrium is reached in CyS, balancing both direct interactions and evolving social factors. This equilibrium is crucial for maintaining stability and adaptability in complex environments, allowing the system to provide relevant recommendations even as the network evolves.

With the exponential rise in online platforms for goods, services, and social networking, users often encounter overwhelming amounts of content, making effective filtering essential (Panzer and Gronau 2024; Liang et al. 2023). Recommendation systems have thus become a staple on such platforms, and the proposed model offers a robust, node-centered approach to meet these demands. By adapting to preferences and the nature of connections within social graphs-representing friendships, trust links, or shared tastes-CyS delivers dynamic, adaptable recommendations across a range of applications (Heshmati et al. 2025).

## 1.1 Formalization of the social graph structure

A social graph is defined as $G = (U, I, E)$, where: 1. $U = \{u_1, u_2, \ldots, u_n\}$ represents the set of nodes in the graph, where each node corresponds to an individual or entity. 2. $I = \{i_1, i_2, \ldots, i_m\}$ denotes the set of items, representing content, places, or products associated with nodes. 3. $E$ is the set of edges, which includes three key types of relationships: (a) Node-to-node connections, $E_U \subseteq U \times U$, representing social links between individuals. (b) Item-to-item relationships, $E_I \subseteq I \times I$, capturing similarities or associations between items. (c) Node-to-item interactions, $E_{UI} \subseteq U \times I$, reflecting interactions between individuals and items.

At a given time $t$, the collective preferences of all nodes are represented as $P(t) = [P_1(t), P_2(t), \ldots, P_N(t)]^T$, where $P_u(t)$ denotes the preference distribution of node $u$.

Social recommendation systems leverage social graphs to analyze user behavior and content preferences by considering both node ($U$) and item ($I$) interactions. These interactions are captured in an interaction matrix $R \in \mathbb{R}^{|U| \times |I|}$, where $R_{ui}$ represents the interaction score between node $u$ and item $i$. This matrix helps infer node preferences ($P_u$) and accounts for the influence of a node's social connections ($v \in N(u)$) on its choices. The proposed model integrates both node-item interactions and hidden node-node relationships to enhance recommendation accuracy.

## 1.2 Challenges of social recommendation systems

1. *Cold-start problem*: New nodes with no recorded interactions (denoted as $C_{\text{new}}(u) = 1$ when $|R_{u,i}| = 0$) lack sufficient data for recommendations. Previous methods, relying on implicit neighbors based on social structures, often choose irrelevant neighbors, reducing recommendation quality (Hwang 2024).
2. *Data sparsity*: node preference data is typically sparse, with far fewer interactions than total possible ones, expressed by the density $\text{Density}(R) = \frac{|R_{ui} > 0|}{|U| \times |I|}$. This sparsity, combined with evolving preferences, limits models in generating diverse and relevant recommendations, challenging scalability as platforms expand.
3. *Neighbor aggregation*: node preferences $P_u$ are influenced by social neighbors $N(u)$. Common approaches aggregate neighbors' traits evenly, creating $P_{u_{new}} = \frac{1}{|N(u)|} \sum_{v \in N(u)} P_v$ for new nodes. However, this deterministic approach lacks flexibility, failing to capture preference variation across social connections.

4. *Social dynamics*: Interactions such as likes and shares offer implicit node feedback, supplementing models when explicit data is unavailable. The presented framework enhances recommendation relevance by addressing traditional models' limitations, including Matrix Factorization MF and Deep Matrix Factorization DMF, in handling sparse data, scalability, and dynamic social cues. It leverages node interconnectedness to enhance recommendation accuracy by identifying preference overlaps among nodes with similar traits. By analyzing profiles and interactions, the model utilizes behavioral similarity within networks, effectively extracting preferences even without direct feedback.

### 1.3 Contributions

This work introduces several key advancements in recommendation systems, showcasing CyS Algorithm. This model blends swarm intelligence, Graph Neural Networks (GNNs), and multi-objective optimization. It is flexible and works across many applications, not just social networks. The main contributions of this work are listed below.

1. *Centrality-driven preference aggregation mechanism*: This mechanism incorporates network centrality metrics-degree, closeness, and betweenness-into node embedding, enabling the model to prioritize influential nodes within the social graph. Each node's impact in the recommendation process is dynamically weighted by centrality-based social influence, enhancing the effect of socially significant nodes.
2. *Hybrid cold-start mitigation via social signals and latent embeddings*: We address the cold-start problem by fusing social signal propagation with latent embedding methods. For nodes without prior interactions ($C_{new}(u)$), the algorithm relies on implicit social signals from neighboring nodes in the graph.
3. *Scalable message-passing framework with parallelization capabilities*: The methodology introduces a scalable GNN-based Message-Passing framework, inherently parallelizable for large-scale networks. Designed for distributed computing, the decentralized nature of swarm intelligence ensures robustness, allowing it to manage large graphs.
4. *Integration of temporal dynamics for adaptive node preferences*: Temporal dynamics are incorporated to adapt node preferences over time. CyS models these preferences as a time-dependent process. This feature allows the algorithm to update recommendations continually based on recent behaviors, enhancing responsiveness to preference shifts.
5. *Theoretical convergence of the preference update process*: A theoretical convergence proof is provided for CyS's preference update process, ensuring stability in the GNN-based Message Passing. The update function is shown in Eq. 10. As $t \to \infty$, the update process converges to a stable fixed point, ensuring reliable and consensus-driven node preferences through convex optimization and spectral graph theory.
6. *Versatility across domains*: The approach demonstrates broad adaptability, effectively applying to diverse domains and underscoring its versatility across recommendation contexts.

7. *Scalability is essential for recommendation systems in large-scale networks*. The methodology addresses this by leveraging parallel processing, enabling efficient handling of massive social graphs with many nodes and items.

### 1.4 Paper organization

The coming sections of this paper are organized as follows: In Sect. 2, We briefly present the related work, Also we identify key methods in this field and highlight their limitations in this field. Section 3 provides the preliminaries, the background, and the assumptions of the proposed approach. Section 4 provides an overview of the problem definition and the proposed methodology. In Section 5, we present the experiment settings and report the results of our experiments. Section 6 shows the ablation study of the model. Section 7 provides concluding remarks and outlines our plans for future work.

## 2 Related work

In recent years, recommendation systems in social networks have advanced considerably. Approaches such as graph-based models, hybrid optimization strategies, and Deep Neural Networks have enhanced their accuracy, scalability, and flexibility. However, social networks remain complex, with diverse interactions, shifting node preferences, and massive datasets that call for even more advanced solutions. The proposed algorithm takes these advancements further by blending concepts from graph theory, swarm intelligence, hypergraph modeling, and multi-objective optimization. This section will explore related work in the field and show where CyS fits into the ongoing evolution of recommendation systems.

### 2.1 Graph-based recommendation systems

Graph-based models have become a key part of modern recommendation systems. In the early days, these systems mostly used Collaborative Filtering (CF) techniques (Chang et al. 2024; Liu et al. 2022), which treated recommendations as a matrix completion task[5]. The goal was to predict missing interactions between nodes and items by breaking the data down into simpler, low-dimensional spaces. Singular Value Decomposition (SVD) and Matrix Factorization (MF) (Zhou et al. 2024; Chang et al. 2024) have been foundational. However, they struggle with the sparse data often found in social networks. More importantly, they couldn't fully capture the intricate connections and relationships between nodes and items that are naturally present in a graph-based setting, limiting their effectiveness in these environments. Despite advances with Graph Convolutional Networks (GCNs) (Lu et al. 2024; Anjiri et al. 2024), they struggle to capture higher-order interactions in social networks. These interactions often involve multiple nodes influencing a single item or group-based preferences where collaborative consumption occurs. GCNs typically model pairwise interactions, which oversimplifies the complexity of social relationships, where node behavior is shaped by multi-way interactions. To address this, researchers have turned to hypergraph structures. Hypergraphs differ from traditional graphs by using hyperedges to connect multiple nodes. This ability makes them particularly well-suited for modeling complex, multi-relational data.

As social networks grow, new nodes are constantly joining to interact, communicate, and entertain. These nodes often don't have any previous interactions, which makes it hard for traditional recommendation systems to give accurate suggestions. Theories like social homophily (Meydani et al. 2024) and social influence suggest that using social relationships can help solve this problem. Social recommendation methods (Chang et al. 2024; Liu et al. 2022; He et al. 2024) combine both social connections between nodes and their interactions with items, helping to tackle the problem of sparse data and make results more accurate.

### 2.2 Semi-supervised and multi-label learning

Semi-supervised learning is critical for recommendation systems with limited labeled data. Sheikhpour et al.'s HSDAFS (Sheikhpour et al. 2025a) uses hypergraph Laplacians to integrate discriminative information from labeled data with the geometrical structure of unlabeled data, ensuring row-sparsity in the projection matrix for feature selection. Similarly, Sheikhpour et al. (2025b) propose a robust semi-supervised multi-label feature selection method (SMFS) that combines shared subspace learning and graph Laplacian-based manifold learning to capture label correlations and preserve data geometry, using $\ell_1$-norm minimization for robustness to outliers. While both methods excel in static settings, without mdeling temporal dynamics or hierarchical structures. Our CyS algorithm addresses the challenge of modeling dynamic, influence-driven preferences in evolving social networks. By integrating node to vector, time-varying centrality, and swarm-like dynamics. It capture both local and global influence in dynamic social graphs.

### 2.3 Spectral methods and graph structure learning

Spectral methods are pivotal in machine learning for clustering, representation learning, and dimensionality reduction. It leverages graph spectral properties to capture complex data relationships. Berahmand et al. (2025) provide a comprehensive survey on spectral clustering with graph structure learning, reviewing a set of techniques that enhance clustering performance. These methods often rely on the eigen spectrum of graph Laplacian matrices to uncover data relationships, typically applied to static or semi-static graphs, such as hyperspectral images (Wang et al. 2019a). Spectral embedding, a prominent dimensionality reduction technique, uses Laplacian eigenvalues to reveal hidden manifold structures, preserving local geometric characteristics and capturing non-linear relationships that traditional linear methods miss (Ding et al. 2024). By representing data points as nodes in a graph, spectral embedding facilitates dimensionality reduction while maintaining topological connections, making it effective for high-dimensional data.

Spectral clustering, a key application of spectral methods, excels at identifying non-convex clusters and capturing non-linear structures through graph-based embeddings. By partitioning data based on connectivity and relationships, spectral clustering is versatile across domains, particularly when combined with robust graph construction techniques. For large-scale datasets, anchor graph-based spectral clustering enhances scalability by using a subset of representative points (anchors) to approximate relationships efficiently, preserving clustering quality while reducing computational costs (Berahmand et al. 2025). However, these methods are less effective for dynamic social networks with evolving topologies, as they rely on fixed graph structures. Additionally, spectral clustering for community detection faces challenges such as high dimensionality, noise sensitivity, complex graph construction, and limited adaptability to dynamic networks (Ding et al. 2024). In contrast, our CyS algorithm adapts spectral analysis to dynamic hypergraphs, utilizing a time-varying propagation matrix $W(t)$ to model preference evolution based on centrality-weighted interactions. This dynamic approach, rooted in spectral graph theory, integrates Node2Vec embeddings and attention-based message-passing to effectively capture evolving influence patterns.

### 2.4 Deep learning and attention mechanisms in recommendation systems

Deep learning models have been effectively applied to recommendation systems, with Multilayer Perceptrons (MLPs) excelling in approximating diverse functions (Sheikh et al. 2019). Social Fusion Network (SFNet) (Zhang et al. 2022), an end-to-end deep model, personalizes recommendations by mapping item and node attributes into a shared space and using skip-connections to capture non-linear relationships. The Wide&Deep model (Cheng et al. 2016) combines linear models for memorization and deep models for generalization, maximizing strengths in both areas. Other advancements include a Reliability-Enhanced Deep Recommender (Bobadilla et al. 2020) that boosts recall, a model (Molaei et al. 2021) leveraging learned social traits for Collaborative Filtering, and a Deep Collaborative model (Behera and Nain 2022) embedding item metadata to address nonlinearity and data sparsity, confirmed effective through embedding vector propagation for missing data.

Attention mechanisms, widely utilized in fields like image processing and computer vision (Zhao et al. 2024; Hassanin et al. 2024), improve focus on critical features by weighting important information. In recommendation systems, this mechanism enhances accuracy by dynamically weighting features based on users' histories and current preferences. Graph Attention Networks (GATs) leverage self-attention to calculate attention coefficients in graphs, enhancing node feature aggregation by representing data as interconnected nodes and edges. Convolution, pooling, and self-attention layers further improve node representations by incorporating neighboring node information. Traditional models often struggle with complex, multi-behavioral data interactions. Our CyS model addresses this by integrating GNN weights to model multi-behavior relationships.

### 2.5 Swarm intelligence and its evolution in recommendation systems

Swarm intelligence, inspired by natural behaviors like bird flocking, ant colonies, and fish schooling, plays a crucial role in solving complex problems, especially in dynamic environments like social networks. Notable techniques in this field include Particle Swarm Optimization (PSO) and Ant Colony Optimization (ACO) (Cai et al. 2021; Awadallah et al. 2024), both of which leverage collective decision-making for optimization. In PSO, the position of each particle in the solution space is updated based on its own experience and the experiences of its neighbors. This process is shown in the velocity update rule in Eq. 1.

$$v_i(t+1) = uv_i(t) + c_1 r_1 (p_i(t) - x_i(t)) + c_2 r_2 (g(t) - x_i(t)) \qquad (1)$$

In this equation, $v_i(t)$ represents the velocity of particle $i$, $p_i(t)$ is its personal best position, and $g(t)$ represents the global best position. PSO balances exploration and exploitation, making it well-suited for continuous optimization. ACO, on the other hand, models optimization as a path-finding process where ants deposit pheromones to guide others. The pheromone update rule is described by Eq. 2:

$$\tau_{ij}(t+1) = (1-\rho)\tau_{ij}(t) + \sum_{k} \Delta\tau_{ij}^{k} \tag{2}$$

here, $\tau_{ij}(t)$ represents the pheromone level on edge $(i, j)$, $\rho$ is the evaporation rate, and $\Delta\tau_{ij}^{k}$ is the pheromone contribution from ant $k$. ACO is particularly effective for discrete tasks like routing. However, its complexity makes it difficult to directly apply in social networks.

The CyS algorithm adapts swarm intelligence specifically for social recommendation. Unlike the traditional models, it dynamically adjusts based on nodes' connectivity and influence, factoring in centrality measures to weight influential nodes more heavily in recommendation outcomes. This centrality-driven approach allows our model to deliver more personalized, intelligent recommendations in fast-changing social networks. Here, each node's influence is continually adjusted according to its centrality, enabling adaptive responses to shifts in node behavior and importance. This approach quantifies each node's social role, ensuring that recommendations are influenced by the most relevant and impactful connections within the network.

## 3 Preliminaries

### 3.1 Background

Inspiration is a cornerstone of both natural and artificial systems, spurring innovation and effective problem-solving by drawing on diverse influences. In Artificial Intelligence (AI) and Machine Learning (ML), this frequently involves looking to nature or social systems for inspiration, fostering algorithms capable of managing complex tasks. To address the limitations of existing recommendation systems, we propose our CyS algorithm, inspired by the interactive dynamics within cyber communities. Our approach, grounded in the dynamics of digital communities.

### 3.2 Assumptions

1. *Dynamic Node Features*: Each node $u_i \in V$ has a feature vector $f_i \in \mathbb{R}^d$, initialized as the concatenated Node2Vec embedding and centrality scores ($f_i = M_i = F_{\text{concat}}(\mathcal{V}_i, [C_{\text{close}}(i), D(i), B(i)])$). Features are updated via interactions in the hypergraph, modeled by:

$$f_i^{(l+1)} = \sigma\left(f_i^{(l)} + \alpha \sum_{u_j \in \mathcal{N}(u_i)} \alpha_{ij}^{(l)} f_j^{(l)}\right) \tag{3}$$

This equation, adapted from social influence theory, governs the message-passing layer (GAT), where $\alpha_{ij}^{(l)}$ are attention coefficients and $\alpha$ is a scaling factor. It ensures node features evolve based on influential neighbors, driving preference propagation.

2. *Preference Convergence*: Sustained interactions in co-preference hyperedges lead to converging preferences, per theorem 1. Nodes $u_i$ and $u_j$ in a hyperedge $e \in E$ (cosine similarity $S_{ij} = \cos(f_i, f_j) > \gamma$) align preferences over iterations:

$$\lim_{l \to \infty} f_i^{(l)} \approx f_j^{(l)} \quad \text{if} \quad u_i, u_j \in e \text{ and } \sum_l \alpha_{ij}^{(l)} > \theta \tag{4}$$

where $\theta$ is interaction threshold. This assumption shapes co-preference hyperedge formation, where the similarity threshold $\gamma$ and sustained message-passing iterations enforce homophily-driven convergence.

3. *Sharing Among Similar Nodes*: Nodes with high cosine similarity ($S_{ij} > \gamma$, where $\gamma$ is the similarity threshold.) share preferences, forming co-preference hyperedges:

$$S_{ij} = \frac{f_i \cdot f_j}{\|f_i\| \|f_j\|} \geq \gamma \implies u_i, u_j \in e \tag{5}$$

Rooted in homophily, this assumption drives hypergraph construction, where similar nodes (based on $f_i$) are grouped, enhancing recommendation scoring ($\mathcal{S}(u, i) = \sum_{v \in \mathcal{N}_u} \mathcal{S}(u, v) \times \mathcal{R}(v, i)$) by leveraging shared preferences.

4. *Selective Sharing Based on Ratings*: Nodes share only positively rated items ($r_{ui} \geq t$) during recommendation:

$$r_{ui} < t \implies \mathcal{R}(u, i) = 0 \tag{6}$$

This assumption, inspired by social filtering, is implemented in preprocessing (threshold $\mathcal{F}_t(R)$) and scoring, ensuring high-quality recommendations, as validated by ablation studies showing improved HR@10 with rating thresholds.

5. *Impact of node interactions*: Nodes with higher interaction degrees $\deg(u_i)$ exert and receive greater influence, with updates given in Eq. 7. Degree centrality highlights influential nodes in the network.

$$\deg(u_i) = \sum_j S_{ij} \quad \text{and} \quad \frac{\partial f_i}{\partial t} = \sum_{j \in \mathcal{N}(u_i)} w_{ij} f_j \tag{7}$$

nodes with higher degrees $\deg(u_i)$ are more central and thus exert and receive more influence. The degree centrality metric $\deg(u_i)$ and the rate of change in node features $\frac{\partial f_i}{\partial t}$ encapsulate the dynamic influence within the network, highlighting key influencers and their impact on node behavior.

6. *Perception and judgment*: Nodes adjust attitudes $a_i$ based on perceptions and judgments, especially when information aligns with existing beliefs. This captures social influence dynamics.

$$a_i^{(t+1)} = a_i^{(t)} + \alpha \cdot \sum_{j \in \mathcal{N}(u_i)} w_{ij}(a_j - a_i) \tag{8}$$

Attitude adjustment through social interaction is modeled by $a_i$, representing the attitude vector of node $i$. The adjustment factor $\alpha$ and the weighted influence $w_{ij}$ encapsulate how nodes modify their attitudes based on their perceptions and judgments of neighboring nodes' messages, reflecting principles of social influence and conformity.

7. *Social Influence and Consistency*: Nodes adjust preferences based on social influence and maintain consistency between preferences and actions, per social judgment and cognitive dissonance theories. This is modeled implicitly through iterative feature updates and recommendation scoring, where preferences ($f_i$) align with shared items ($\mathcal{R}(u,i)$).

## 4 Methodology

The CyberSwarm algorithm is a recommendation system inspired by cyber community dynamics, modeling social networks as nodes connected by edges that represent interdependencies like friendship, shared interests, or transactions. Our proposed algorithm incorporates principles from social psychology to model attitude formation and evolution within social networks. Key theories include: *Social-Judgment Theory*, which describes attitude adjustments based on message evaluation and trusted sources. *Consistency Theory*, highlighting individuals' alignment of behaviors to maintain cognitive consistency. *Self-Perception Theory*, where attitudes emerge from observing personal behaviors. *Functional Theory*, emphasizing attitudes as tools for identity expression or group adaptation.

We designed CyS to capture social dynamics in node interactions for improved recommendations. It constructs a hypergraph $\mathcal{H}$ to represent these relationships. The algorithm contains the following main steps: (1) Generating a feature map $M_i$ using centrality measures $\mathcal{M}_{\text{central}}(\mathcal{H})$ and Node2Vec algorithm $F_{\text{node2vec}}$. (2) A learning method updates this map based on social interactions, simulating social dynamics. (3) The similarity $\mathcal{S}(u,i)$ between nodes is measured to identify the most similar ones. (4) Similar nodes' preferences are recommended to each other. The following sections detail these steps as shown in Fig. 1.

**Theorem 1** (Dynamic Collaborative Swarm Equilibrium (DCSE)) Let $\mathcal{G} = (V, E)$ be weighted social graph with a dynamic centrality-weighted propagation matrix and $W(t) \in \mathbb{R}^{|V| \times |V|}$. The node preferences $P(t)$ converge to a time-varying equilibrium $P^\infty(t) = \nu(t)$, the Perron-Frobenius eigenvector of $W(t)$.

The algorithm updates preferences $P_i(t)$ using time-varying centrality metrics, unlike Graph Neural Networks' (GNNs) static aggregation. The centrality weight between nodes $i$ and $j$ is:

$$a_{ij}(t) = \lambda_1 D_{ij}(t) + \lambda_2 C_{ij}(t) + \lambda_3 B_{ij}(t), \tag{9}$$

where $D_{ij}(t)$, $C_{ij}(t)$, and $B_{ij}(t)$ are degree, closeness, and betweenness centralities, and $\lambda_1, \lambda_2, \lambda_3$ are hyperparameters. The preference update is:

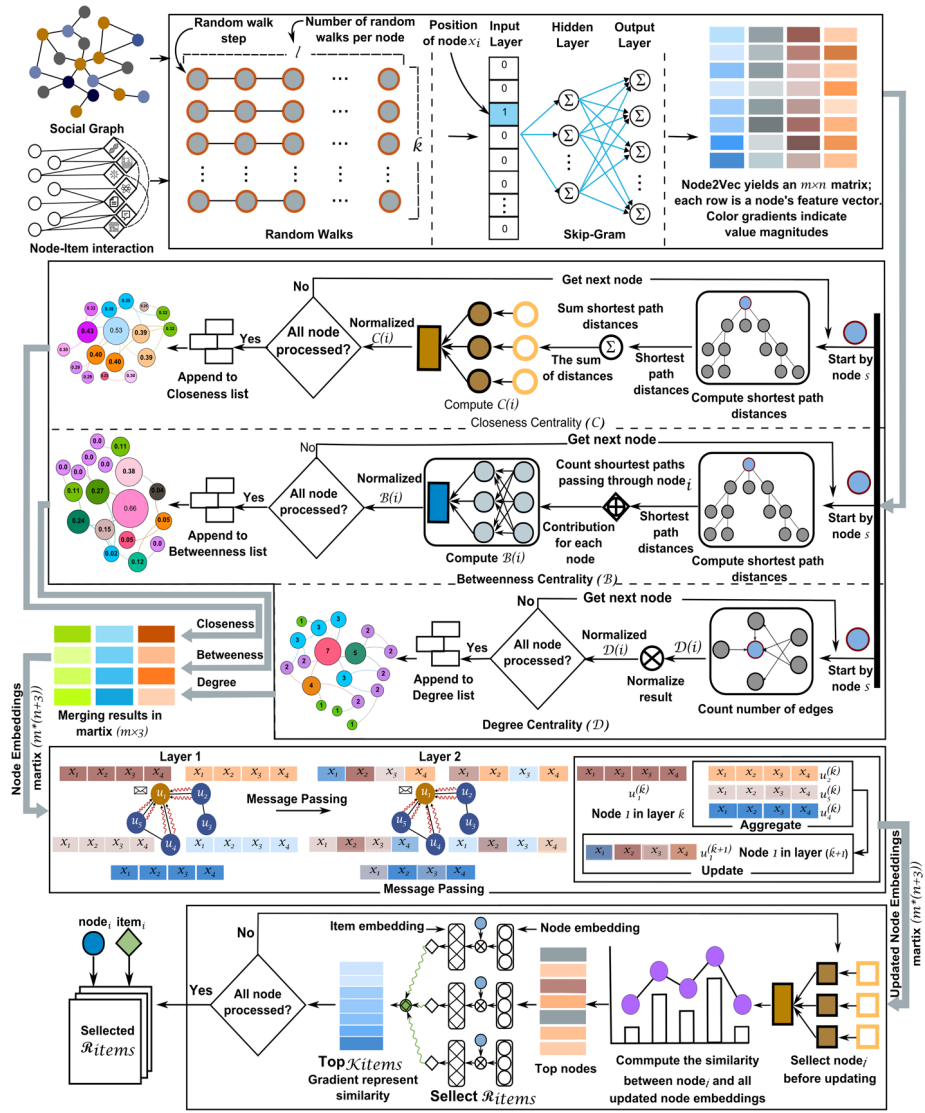

**Fig. 1** General workflow of the CyberSwarm algorithm

$$P_i(t+1) = (1-\eta)P_i(t) + \eta \sum_{j \in N(u_i)} a_{ij}(t)M(P_j(t)), \qquad (10)$$

where $0 < \eta \leq 1$ is the learning rate, and $M(P_j(t))$ is node $j$'s centrality-weighted influence. The propagation matrix $W(t) \in \mathbb{R}^{N \times N}$ is:

$$W_{ij}(t) = \begin{cases} a_{ij}(t), & \text{if } j \in N(u_i, t), \\ 1 - \eta, & \text{if } i = j, \\ 0, & \text{otherwise,} \end{cases} \qquad (11)$$

with $\sum_{j \in N(u_i)} a_{ij}(t) = 1$, ensuring row-stochasticity ($\sum_j W_{ij}(t) = 1$). The network evolves as:

$$P(t+1) = W(t)P(t), \qquad (12)$$

yielding:

$$P(t) = W(t-1)W(t-2)\ldots W(0)P(0). \qquad (13)$$

**Lemma 1** (Primitivity of the Propagation Matrix) If the social graph $\mathcal{G} = (V, E)$ is strongly connected at each time $t$, and the centrality weights $a_{ij}(t) > 0$ for all $j \in N(u_i, t)$, then the propagation matrix $W(t)$ is primitive.

**Proof** Since $\mathcal{G}$ is strongly connected, for any nodes $u_i, u_j \in V$, there exists a path $u_i = v_0 \to v_1 \to \cdots \to v_k = u_j$ of length at most $k \leq |V|$. For each edge $(v_m, v_{m+1})$ in the path, $W_{v_m v_{m+1}}(t) = a_{v_m v_{m+1}}(t) > 0$ by the lemma's assumption. Additionally, $W_{ii}(t) = 1 - \eta > 0$ for all $i$. Consider the matrix power $W(t)^k$. The $(i,j)$-entry of $W(t)^k$ is positive if there exists a sequence of nodes $v_0 = u_i, v_1, \ldots, v_k = u_j$ such that $\prod_{m=0}^{k-1} W_{v_m v_{m+1}}(t) > 0$. Since strong connectivity ensures such a path exists, and all relevant entries are positive, $[W(t)^k]_{ij} > 0$ for all $i, j$. Thus, $W(t)^k > 0$, so $W(t)$ is primitive. □

**Lemma 2** (Convergence in Non-Strongly Connected Graphs) If the social graph $\mathcal{G} = (V, E)$ at time $t$ consists of weakly connected components $C_1, \ldots, C_m$, each with an aperiodic induced subgraph, then within each component $C_r$, preferences $P_i(t)$ for $v_i \in C_r$ converge to a local equilibrium $\nu_r(t)$, the Perron-Frobenius eigenvector of the submatrix $W_r(t)$ restricted to $C_r$.

**Proof** For each weakly connected component $C_r$, the submatrix $W_r(t)$ is row-stochastic, as $\sum_{j \in N(u_i) \cap C_r} a_{ij}(t) = 1$ for $v_i \in C_r$. Since $C_r$'s induced subgraph is aperiodic, $W_r(t)$ is primitive (as aperiodicity ensures some power has positive entries along paths). By Perron-Frobenius, $W_r(t)$ has a unique eigenvalue $\lambda_1^{(r)}(t) = 1$ with a positive eigenvector $\nu_r(t)$. Preferences within $C_r$ evolve as $P_r(t+1) = W_r(t)P_r(t)$. Using spectral decomposition, non-dominant eigenvalues $|\lambda_i^{(r)}(t)| < 1$ decay, so $P_r(t) \to \nu_r(t)$ as $t \to \infty$, forming a local equilibrium. □

**Proof of Theorem 1** Assume the social graph is strongly connected, so $W(t)$ is row-stochastic and primitive by Lemma 1. By the Perron-Frobenius theorem, $W(t)$ has a unique dominant eigenvalue $\lambda_1(t) = 1$ with a positive right eigenvector $\nu(t)$, normalized such that $\|\nu(t)\|_1 = 1$. Other eigenvalues satisfy $|\lambda_i(t)| < 1$ for $i \geq 2$.

The preference evolves as $P(t) = \prod_{k=0}^{t-1} W(k)P(0)$ (Eq. 20). Assume $W(t)$ evolves smoothly, i.e., $\|W(t) - W(t-1)\|_2$ is small, ensuring $\nu(t)$ changes continuously. Define the product matrix $W_{\text{prod}}(t) = \prod_{k=0}^{t-1} W(k)$, which is row-stochastic.

Express $P(t)$ in the time-varying eigenbasis of $W(t-1)$.

Let $\{v_1(t-1), v_2(t-1), \ldots, v_N(t-1)\}$ be the right eigenvectors of $W(t-1)$, with $v_1(t-1) = \nu(t-1)$. At time $t-1$:

$$P(t-1) = \sum_{i=1}^{N} \alpha_i(t-1)v_i(t-1), \tag{14}$$

where $\alpha_i(t-1)$ are coefficients. Applying $W(t-1)$:

$$P(t) = W(t-1)P(t-1) = \sum_{i=1}^{N} \alpha_i(t-1)\lambda_i(t-1)v_i(t-1). \tag{15}$$

Since $\lambda_1(t-1) = 1$ and $|\lambda_i(t-1)| < 1$ for $i \geq 2$, non-dominant components decay. Over iterations, as $t \to \infty$, the product $W_{\text{prod}}(t)$ amplifies the component along $\nu(t)$. Given $W(t)$'s primitivity and smooth variation, $W_{\text{prod}}(t)$ converges to a rank-1 projection onto $\nu(t)$, so:

$$\lim_{t \to \infty} \frac{P(t)}{\|P(t)\|_1} = \nu(t). \tag{16}$$

Since $P(t)$ remains normalized ($\sum_i P_i(t) = 1$) due to row-stochasticity, $P(t) \to \nu(t)$.

Unlike GNNs' static local aggregation, $W(t)$'s dynamic, centrality-driven structure ensures global consensus, addressing the need for a distinct model. This proves convergence to $P^\infty(t) = \nu(t)$, satisfying the theorem with rigorous spectral grounding. □

**Theorem 2** (CyberSwarm Emergent Hierarchical Synchronization (CEHS)) Let $\mathcal{G} = (V, E, L)$ be a multi-layered hierarchical graph with layers $L = \{L_1, L_2, \ldots, L_k\}$, and let $W_H(t), W_V(t) \in \mathbb{R}^{|V| \times |V|}$ be dynamic propagation matrices for horizontal and vertical influence, respectively. The node preferences $P(t)$ converge to a time-varying hierarchical equilibrium, synchronizing within and across layers, governed by the spectral properties of $W_H(t)$ and $W_V(t)$.

The CyS algorithm extends the Dynamic Collaborative Swarm Equilibrium (DCSE) by modeling synchronization in hierarchically structured social networks, unlike Graph Neural Networks (GNNs) that assume flat structures. Node preferences $P_i(t)$ are updated by integrating intra-layer (horizontal) and inter-layer (vertical) influences:

$$P_i(t+1) = (1-\eta)P_i(t) + \eta \left( \sum_{v_j \in N_H(v_i)} \alpha_{ij}(t) M(P_j(t)) + \sum_{v_j \in N_V(v_i)} \beta_i(t) N(P_j(t)) \right), \tag{17}$$

where $0 < \eta \leq 1$ is the learning rate, $N_H(v_i)$ and $N_V(v_i)$ denote horizontal and vertical neighbors, $\alpha_{ij}(t)$ and $\beta_i(t)$ are time-varying centrality-based weights, and $M(\cdot)$, $N(\cdot)$ are transformation functions for intra- and inter-layer influence, respectively.

The system evolves via horizontal and vertical propagation matrices $W_H(t)$ and $W_V(t)$, defined as:

$$[W_H(t)]_{ij} = \begin{cases} \eta \alpha_{ij}(t), & \text{if } v_j \in N_H(v_i), \\ 1 - \eta, & \text{if } i = j, \\ 0, & \text{otherwise,} \end{cases}$$
$$[W_V(t)]_{ij} = \begin{cases} \eta \beta_i(t), & \text{if } v_j \in N_V(v_i), \\ 0, & \text{otherwise,} \end{cases} \quad (18)$$

with $\sum_{j \in N_H(v_i)} \alpha_{ij}(t) = 1$, ensuring $W_H(t)$ is row-stochastic ($\sum_j [W_H(t)]_{ij} = 1$). The combined update is:

$$P(t+1) = [W_H(t) + W_V(t)]P(t) = W_C(t)P(t), \quad (19)$$

yielding:

$$P(t) = \prod_{s=0}^{t-1} W_C(s) P(0). \quad (20)$$

**Lemma 3** (Primitivity of Horizontal Propagation Matrix) If the subgraph induced by layer $L_\ell$ in $\mathcal{G} = (V, E, L)$ is strongly connected at each time $t$, and the centrality weights $\alpha_{ij}(t) > 0$ for all $v_j \in N_H(v_i)$, then the restricted horizontal propagation matrix $W_H^{(\ell)}(t) = \Pi_\ell W_H(t) \Pi_\ell$ is primitive.

**Proof** Since the subgraph induced by $L_\ell$ is strongly connected, for any nodes $v_i, v_j \in L_\ell$, there exists a path $v_i = u_0 \to u_1 \to \cdots \to u_k = v_j$ of length at most $k \leq |L_\ell|$. For each edge $(u_m, u_{m+1})$ in $L_\ell$, $[W_H(t)]_{u_m u_{m+1}} = \eta \alpha_{u_m u_{m+1}}(t) > 0$, and $[W_H(t)]_{ii} = 1 - \eta > 0$. The projection $\Pi_\ell$ ensures $[W_H^{(\ell)}(t)]_{ij} = [W_H(t)]_{ij}$ for $v_i, v_j \in L_\ell$, and 0 otherwise. For $[W_H^{(\ell)}(t)^k]_{ij}$, a positive entry requires a path in $L_\ell$ such that $\prod_{m=0}^{k-1} [W_H^{(\ell)}(t)]_{u_m u_{m+1}} > 0$. Strong connectivity and $\alpha_{ij}(t) > 0$ ensure such a path exists, making $[W_H^{(\ell)}(t)^k]_{ij} > 0$ for all $i, j \in L_\ell$. Thus, $W_H^{(\ell)}(t)^k > 0$, so $W_H^{(\ell)}(t)$ is primitive. $\square$

**Proof of Theorem 2** Assume $\mathcal{G}$ is have a $W_H(t)$ row-stochastic and primitive within each layer $L_\ell$. Assume $W_V(t)$ ensures vertical influence propagation, and both matrices evolve smoothly, i.e., $\|W_H(t) - W_H(t-1)\|_2$ and $\|W_V(t) - W_V(t-1)\|_2$ are small. The combined matrix $W_C(t) = W_H(t) + W_V(t)$ governs preference evolution.

For each layer $L_\ell$, restrict $W_H(t)$ to $W_H^{(\ell)}(t) = \Pi_\ell W_H(t) \Pi_\ell$, where $\Pi_\ell$ is the projection onto $L_\ell$:

$$[\Pi_\ell]_{ij} = \begin{cases} 1, & \text{if } v_i, v_j \in L_\ell, \\ 0, & \text{otherwise.} \end{cases} \quad (21)$$

By Perron-Frobenius, $W_H^{(\ell)}(t)$ has a dominant eigenvalue $\lambda_1^{(\ell)}(t) = 1$ with eigenvector $\nu_\ell(t) > 0$, and $|\lambda_i^{(\ell)}(t)| < 1$ for $i \geq 2$. Thus, preferences within $L_\ell$ synchronize to $\nu_\ell(t)$.

Express the preference vector as:

$$P(t-1) = \sum_{\ell=1}^{k} \sum_{i=1}^{|L_\ell|} \alpha_{i,\ell}(t-1) v_{i,\ell}(t-1), \tag{22}$$

where $\{v_{i,\ell}(t-1)\}$ are eigenvectors of $W_H^{(\ell)}(t-1)$, and $v_{1,\ell}(t-1) = \nu_\ell(t-1)$. Applying $W_C(t-1)$:

$$\begin{aligned} P(t) = W_C(t-1) P(t-1) &= \sum_{\ell=1}^{k} \Pi_\ell W_H(t-1) P(t-1) \\ &+ \sum_{\ell \neq \ell'} \Pi_\ell W_V(t-1) \Pi_{\ell'} P(t-1). \end{aligned} \tag{23}$$

The first term drives intra-layer synchronization, as non-dominant eigenvalues decay. The second term couples layers via vertical influence.

Assume vertical coupling strength $\Gamma_{\ell\ell'}(t) = \max_{v_i \in L_\ell} \sum_{v_j \in L_{\ell'} \cap N_V(v_i)} \beta_i(t)$ is sufficient for inter-layer influence. As $t \to \infty$, the product $\prod_{s=0}^{t-1} W_C(s)$ projects preferences onto a time-varying hierarchical equilibrium, where nodes within each $L_\ell$ synchronize to $\nu_\ell(t)$, and vertical propagation aligns $\nu_\ell(t)$ across layers, forming a global equilibrium linked to DCSE's $\nu(t)$.

Since $W_H(t)$ ensures intra-layer row-stochasticity and $W_V(t)$ propagates vertical influence smoothly, $P(t)$ converges to:

$$\lim_{t \to \infty} \frac{P(t)}{\|P(t)\|_1} = \nu(t), \tag{24}$$

where $\nu(t)$ reflects the hierarchical structure. Unlike GNNs' flat aggregation, CEHS's dual-channel dynamics ensure hierarchical synchronization, proving the theorem with spectral grounding. □

### 4.1 Hyper-graph representation

The hypergraph representation is central to the CyberSwarm (CyS) algorithm, modeling complex social interactions in a hypergraph $\mathcal{H} = (V, E)$, where $V$ represents users and $E$ denotes hyperedges capturing multi-node relationships. Unlike pairwise graphs, hyperedges enable CyS to encode group dynamics, supporting adaptive recommendations. We define hyperedge semantics and modeling assumptions below to clarify their construction.

1. *Ego Network*: A social graph $G = (V, E)$ models pairwise connections. For each node $v \in V$, its ego network $G_v = (V_v, E_v)$ includes $v$, its neighbors $V_v \subseteq V$, and edges $E_v \subseteq E$, informing local relationships for hyperedge formation.
2. *Node-Item Interactions*: User-item interactions form a bipartite graph $H = (U, I, R)$, with users $U \subseteq V$, items $I$, and interactions $R \subseteq U \times I$. The matrix $R \in \mathbb{R}^{|U| \times |I|}$ encodes interaction strength $r_{ui}$ for user $u \in U$ and item $i \in I$, guiding co-interaction hyperedges.
3. *Hyperedge Semantics*: Hyperedges are defined by *co-interaction* and *co-preference* criteria:

   - Co-Preference: A hyperedge groups users with similar preferences, based on Node-2Vec embeddings $\mathcal{V}_i$ combined with the Centrality scores $(D(i), C_{close}(i), B(i))$ as shown in Eq. 33. Users form a hyperedge if their similarity measure are large (e.g., cosine similarity exceed 0.7) and $|e| \geq 2$, addressing sparse or cold-start scenarios.
   - Co-Interaction: A hyperedge $e = \{u \in U \mid r_{ui} \neq 0\}$ groups nodes interacting with item $i \in I$ within a time window $t$, if $|e| \geq 2$. This captures shared engagements, like rating the same item.

Co-interaction hyperedges model group influence, while co-preference hyperedges mitigate sparsity, per Theorem 1, with time windows ensuring temporal relevance.

### 4.2 Pre-processing

To address data variability, irregularities, and potential adversarial behaviors in hypergraphs, we introduce a dedicated preprocessing layer to produce a clean hypergraph $\mathcal{H} = (V, E)$. Given an input graph $G$, the cleaned graph $C$ is obtained through the cleaning function $C \leftarrow \mathcal{P}_{\text{clean}}(G)$, where $\mathcal{P}_{\text{clean}}$ encapsulates operations to enhance node interactions and relationships. These operations include removing duplicate nodes, eliminating isolated nodes, applying degree filtering, enforcing rating thresholds, handling missing values, detecting anomalous nodes, and computing trust scores to mitigate noise and manipulative behaviors.

In the hypergraph $\mathcal{H} = (V, E)$, a node $v \in V$ is classified as isolated if $\deg(v) = 0$, where the degree is defined as $\deg(v) = \sum_{e \in E} \chi(v \in e)$, with $\chi(v \in e) = 1$ if $v \in e$ and 0 otherwise. Isolated nodes introduce noise and computational inefficiencies, so they are removed: $V' = V \setminus \mathcal{V}_{\text{iso}}$, where $\mathcal{V}_{\text{iso}} = \{v \in V \mid \deg(v) = 0\}$.

To filter poorly rated items and ensure high-quality recommendations, we apply a rating threshold. The rating $R(u, i)$ represents the score node $u \in U$ assigns to item $i \in I$, with $R : U \times I \to [0, 5]$. The filtering operator $\mathcal{F}_t$ is defined as:

$$\mathcal{F}_t(R) = \{(u, i) \in U \times I \mid R(u, i) \geq t\} \quad (25)$$

The filtered rating set is $R_{\text{filtered}} = \mathcal{F}_t(R)$, enhancing recommendation relevance and supporting precise feature vector generation as shown in ablation study (Table 9).

To counter adversarial interactions (e.g., fake ratings, sybil attacks), we introduce an anomaly detection filter. Each node $u_i \in U$ is assigned an anomaly score $\psi_i$ based on interaction frequency and rating inconsistency:

$$\psi_i = \frac{|\{r_{ui} \neq 0\}|}{\max_{u_j \in U} |\{r_{uj} \neq 0\}|} \cdot \mathrm{Var}\left(\{r_{ui} \mid i \in I, r_{ui} \neq 0\}\right) \tag{26}$$

Nodes with $\psi_i > \phi$ (e.g., $\phi = 0.9$) are flagged as potential adversaries and excluded from hyperedge formation. This filter, applied within $\mathcal{P}_{\mathrm{clean}}$, mitigates manipulative behaviors, complementing the rating threshold.

To handle conflicting signals during preference updates, we compute trust scores $\tau_{ij} \in [0, 1]$ between nodes $u_i, u_j \in U$, based on rating consistency within a time window:

$$\tau_{ij} = \exp\left(-\mathrm{Var}\left(\{r_{ui} - r_{uj} \mid i \in I, r_{ui}, r_{uj} \neq 0\}\right)\right) \tag{27}$$

Trust scores are used in hyperedge formation and message-passing, ensuring reliable social interactions. These preprocessing steps produce a cleaned graph $C$, enabling robust hypergraph construction and accurate learning from social interactions.

### 4.3 Centrality measures

Centrality measures play a vital role in feature extraction. We propose the use of them to identify influential nodes to guide swarm behavior and decision-making. By evaluating centrality, the model constructs detailed node representations, where central nodes act as hubs, enhancing communication, coordination, and adaptability to changes or threats. The algorithm employs Degree Centrality $D(i)$, Closeness Centrality $C_{\mathrm{close}}(i)$, and Betweenness Centrality $B(i)$, as defined in Eq. 28. These measures offer distinct perspectives on network structure and dynamics, crucial for optimizing swarm responses and strategies.

$$\mathcal{M}_{\mathrm{central}}(\mathcal{H}) = \{C_{\mathrm{close}}(i), D(i), B(i)\} \tag{28}$$

#### 4.3.1 Degree centrality

Degree Centrality is key in the CyS, quantifying each node's direct connections to identify influential nodes. These nodes shape network dynamics and information flow. Incorporating Degree Centrality into feature vectors ensures accurate representation of connected, influential nodes, enhancing the algorithm's ability to model social dynamics and make effective recommendations. For a node $v$, Degree Centrality $D(v)$ is shown in Eq. 29.

$$D(v) = \deg(v) \tag{29}$$

Figure 2 demonstrates node degree centrality within the network. High-degree nodes, prominently positioned at the center, indicate their critical role as influential hubs facilitating information flow. Clusters of smaller nodes, differentiated by color, reveal varying connectivity levels and frequent interactions among nodes with similar dynamics. This highlights the algorithm's ability to identify and emphasize central nodes, which significantly impact the model by prioritizing key influencers in the network.

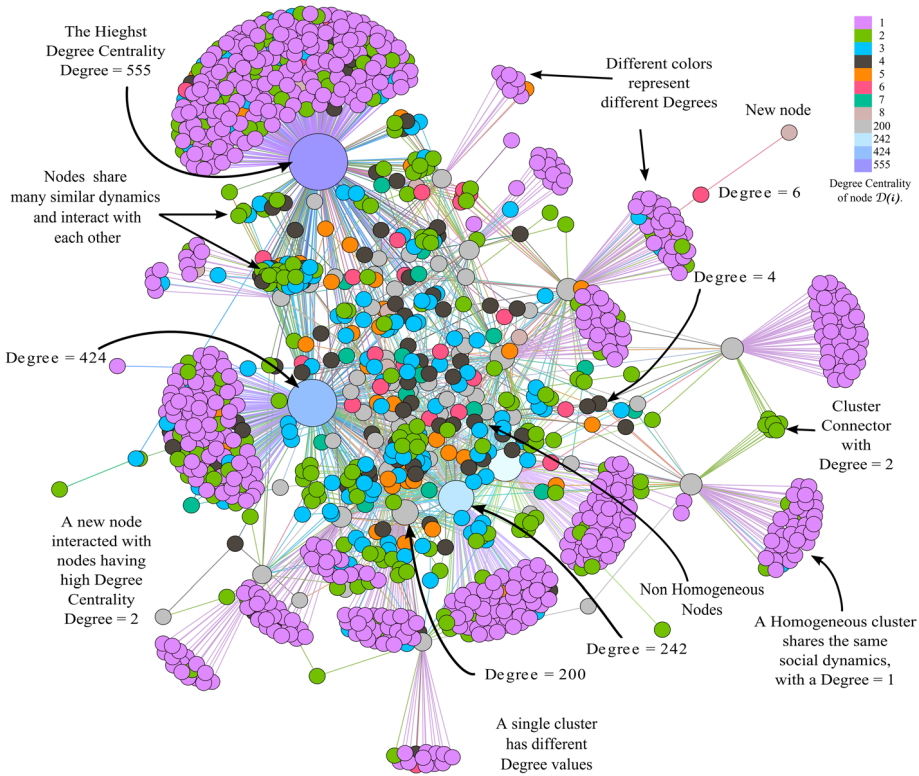

**Fig. 2** This figure illustrates the degree centrality within the network, where the size of each node represents its degree centrality. Clusters of smaller nodes, differentiated by color

### 4.3.2 Closeness centrality

Closeness Centrality is used to measure how close a node is to all other nodes in the network. It is defined as the reciprocal of the sum of the shortest path distances from a node to all other nodes. For a node $v$, Closeness Centrality $C_{close}(v)$ is given by Eq. 30. In the proposed algorithm, nodes with high Closeness Centrality are crucial as they can quickly communicate with other nodes.

$$C_{close}(v) = \frac{1}{\sum_{u \neq v} d(v, u)} \quad (30)$$

Fig. 3 highlights closeness centrality, where larger nodes represent those with higher proximity to all others in the network, marking them as efficient information spreaders. Clusters with similar closeness centrality indicate balanced access to communication pathways. The algorithm leverages this to prioritize nodes that optimize information dissemination, ensuring critical messages quickly reach the entire network.

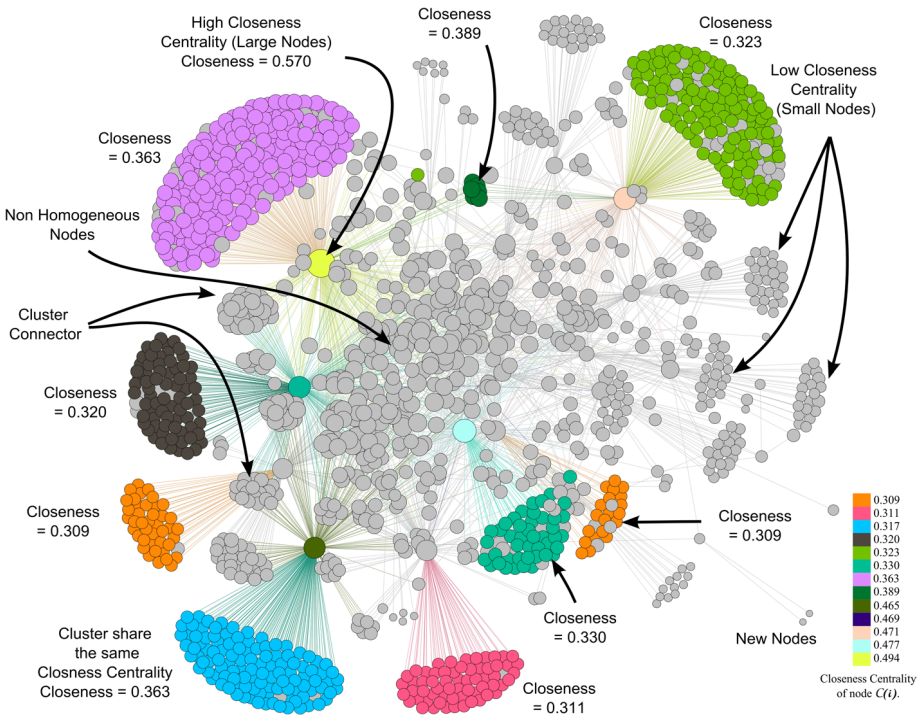

**Fig. 3** This figure shows the closeness centrality of nodes, with larger nodes indicating higher closeness centrality. The CyS's emphasis on these nodes enhances the algorithm's efficiency in spreading information throughout the network

### 4.3.3 Betweenness centrality

Nodes with high Betweenness Centrality are crucial for managing network information flow, acting as intermediaries on shortest paths. Their role enhances communication and influence across the system. Incorporating Betweenness Centrality into feature vectors identifies these pivotal nodes, improving recommendation accuracy and overall algorithm performance. For a node $v$, Betweenness Centrality $C_B(v)$ is computed using Eq. 31.

$$C_B(v) = \sum_{s \neq v \neq t} \frac{\sigma_{st}(v)}{\sigma_{st}} \tag{31}$$

### 4.4 Node2Vec embeddings

The Node2Vec algorithm is integral in our framework as it translates each node into a numerical vector, effectively capturing local and global structures of the network. This transformation is mathematically described in Eq. 32.

$$\mathcal{V}_i \leftarrow F_{\text{node2vec}}\left(v_i, \{\mathcal{W}_k\}_{k=1}^{K}, \{\mathcal{Q}_k\}_{k=1}^{K}\right) \tag{32}$$

where $\mathcal{V}_i$ represents the embedding of node $v_i$, $\mathcal{F}_{\text{node2vec}}$ is the Node2Vec function, $\{\mathcal{W}_k\}_{k=1}^{K}$ are sets of random walks, and $\{\mathcal{Q}_k\}_{k=1}^{K}$ are sets of walk strategies. The key parameters in Node2Vec are the random walk length ($l$) and the number of walks per node ($k$). Random walks of length $l$, starting from node $v_i$, are represented as $\mathcal{W}_k(v_i) = \{w_0, w_1, \ldots, w_{l-1} \mid w_0 = v_i\}$. These walks influence how effectively node embeddings capture the network structure. The shorter walks focus on local, immediate neighbors and longer walks cover more distant connections, as shown in Fig. 1. The parameter $k$ (number of walks per node) enhances embedding robustness, as multiple walks offer a richer representation but increase computational complexity. These parameters influence embedding quality, requiring careful tuning to capture both local and global structures in complex hypergraph data accurately.

The resulting embeddings integrate with centrality measures for each node's feature representation, as shown in Eq. 33. Here, $\mathcal{V}_i$ denotes the Node2Vec embedding for $v_i$, and $[C_{close}(i), D(i), B(i)]$ represent centrality metrics-quantifying influence and connectivity. The concatenation function $F_{\text{concat}}$ combines features into a unified vector $M_i$, where $\mathcal{V}_i \in \mathbb{R}^d$ and $[C_{\text{close}}(i), D(i), B(i)] \in \mathbb{R}^3$ form $M_i \in \mathbb{R}^{d+3}$. This combination may be weighted by $\mathcal{W}_{\text{concat}}$ to balance each component's contribution. This yields a robust, multi-dimensional representation of node $v_i$, merging embeddings and centrality metrics to support the CyS's downstream tasks, as outlined in CyS Algorithm.

$$M_i \leftarrow F_{\text{concat}}\left(\mathcal{V}_i, [C_{\text{close}}(i), D(i), B(i)], \mathcal{W}_{\text{concat}}\right) \tag{33}$$

### 4.5 Message-passing layer

We propose a learning phase to represent the inspiration of the influence process. This layer is the Message Passing layer and contains two phases, Aggregation and Updating.

### 4.5.1 Aggregation

The aggregation phase focuses on capturing influence and information flow among nodes in a network by aggregating neighbor data and updating each node's representation. For any node $i \in V$, with neighbors $\mathcal{N}(i)$, the aggregation function $\mathcal{A}_{\text{agg}}$ at layer $l$ computes the message from node $j \in \mathcal{N}(i)$ to node $i$ using the feature vectors $h_i^{(l)}$ and $h_j^{(l)}$, guided by attention mechanisms. The message $m_{ij}^{(l)}$ is defined in Eq. 34.

$$m_{ij}^{(l)} \leftarrow \sum_{j \in \mathcal{N}(i)} \exp\left(a \cdot \left[h_i^{(l)} \| h_j^{(l)} \| W h_i^{(l)} \| W h_j^{(l)}\right]\right) h_j^{(l)} \tag{34}$$

where $m_{ij}^{(l)}$ represents the message passed from node $j$ to node $i$, $\alpha_{ij}^{(l)}$ is the attention coefficient quantifying the relevance of node $j$ to node $i$, and $\mathcal{W}$ is the transformation matrix applied to the feature vectors. The concatenation operation is denoted by $\|$. To ensure proper weighting, $\alpha_{ij}^{(l)}$ is normalized using the softmax function, as shown in Eq. 35.

$$\alpha_{ij}^{(l)} = \frac{\exp\left(a \cdot \left[h_i^{(l)} \| h_j^{(l)} \| W h_i^{(l)} \| W h_j^{(l)}\right]\right)}{\sum_{k \in \mathcal{N}(i)} \exp\left(a \cdot \left[h_i^{(l)} \| h_k^{(l)} \| W h_i^{(l)} \| W h_k^{(l)}\right]\right)} \tag{35}$$

The aggregated message $m_i^{(l)}$ is the weighted sum of all neighbor messages:

$$m_i^{(l)} = \sum_{j \in \mathcal{N}(i)} \alpha_{ij}^{(l)} h_j^{(l)} \tag{36}$$

This phase enables the model to prioritize relevant information from neighbors, effectively capturing local and global network structures.

### 4.5.2 Updating

The update function refines node features using neighbor information. CyS uses attention to weigh contributions, calculated by the softmax in Eq. 37.

$$\alpha_{i,j}^{(l)} \leftarrow \text{softmax}\left(e^{a \cdot \left[h_i^{(l)} \| h_j^{(l)}\right]} \Big/ \sum_{k \in \mathcal{N}(i)} e^{a \cdot \left[h_i^{(l)} \| h_k^{(l)}\right]}\right) \tag{37}$$

here, $\alpha_{i,j}^{(l)}$ is the normalized attention coefficient based on the concatenated features $h_i^{(l)}$ and $h_j^{(l)}$, and parameter $a$ controls the importance of each neighbor. Equation 38 provides the final updated representation incorporating attention:

$$h_i^{(l+1)} \leftarrow \sigma\left(\sum_{j=1}^{|\mathcal{N}(i)|} \alpha_{i,j}^{(l)} m_{ij}^{(l)} + b_i^{(l)}\right) \tag{38}$$

In this formulation, $\alpha_{i,j}^{(l)}$ weights each message $m_{ij}^{(l)}$, and $b_i^{(l)}$ is a bias term. The parameters $W$, $a$, $\sigma$, and $b_i^{(l)}$ enable transformation, attention, non-linearity, and fine-tuning. By iteratively updating node representations, the model captures dynamic preferences and characteristics, enhancing recommendation accuracy.

### 4.6 Recommendation layer

In this layer, The recommendation process in our algorithm leverages the computed similarity measures to suggest items to nodes based on their interactions and preferences. It contains two main steps, the Similarity measure and the recommendation process.

### 4.6.1 Similarity measure

We propose a measure to compute similarity between a node and others using Euclidean, Jaccard, or Cosine similarity, as shown in Eqs. 39, 40, and 41. These distances help select the most similar node $u$ from each set $\mathcal{U}$.

$$d_{si} \leftarrow \sqrt{\sum_{k=1}^{n}(\mathcal{H}[s][k] - \mathcal{H}[i][k])^2} \tag{39}$$

$$d_{si} \leftarrow \frac{|\mathcal{H}[s] \cap \mathcal{H}[i]|}{|\mathcal{H}[s] \cup \mathcal{H}[i]|} \tag{40}$$

$$d_{si} \leftarrow \frac{\sum_{k=1}^{n} \mathcal{H}[s][k] \cdot \mathcal{H}[i][k]}{\sqrt{\sum_{k=1}^{n}(\mathcal{H}[s][k])^2} \cdot \sqrt{\sum_{k=1}^{n}(\mathcal{H}[i][k])^2}} \tag{41}$$

here, $\mathcal{H}[s][k]$ and $\mathcal{H}[i][k]$ are the feature values of state $s$ and item $i$ for feature $k$, and $n$ is the number of features. The calculated distance $d_{si}$ is then added to the set $\mathcal{D}$.

### 4.6.2 Recommendation process

First we calculate the item scores. For each item $i$ in the dataset, we calculate the score $\mathcal{S}(u,i)$ for node $u$ by summing the weighted interactions $\mathcal{R}(v,i)$ of the node's neighbors $\mathcal{N}_u$ as we propose in Eq. 42

$$\mathcal{S}(u,i) \leftarrow \sum_{v \in \mathcal{N}_u} \mathcal{S}(u,v) \times \mathcal{R}(v,i) \tag{42}$$

This score measures the relevance of item $i$ to node $u$ based on interactions from similar nodes. The top-K items for node $u$ are then selected using Eq. 43, choosing items with the highest scores $\mathcal{S}(u,i)$.

$$\text{TopKItems}(u) \leftarrow \mathcal{T}_{topK}(\mathcal{S}(u,i)) \tag{43}$$

**Require:** $G_v = (V_v, E_v) \quad \forall v \in V$; // Social Connections
**Require:** $H = (U, I, R)$ // Node-Item Interactions, $R \in \mathbb{R}^{|U| \times |I|}$
**Require:** $F$ // Message-passing functions
 1: $C \leftarrow \mathcal{P}_{clean}(G)$ // Reduce noise in the hyper-graph data
 2: $\mathcal{H} \leftarrow \mathcal{P}_{hyper}(C, G)$ // Generate a hypergraph from the given data
 3: $P_{\mathcal{H}} \leftarrow \mathcal{P}_{batch}(\mathcal{H}, N_{batches}, Max_{iter}, \epsilon)$ // Batch Processing of Hyper-graph $\mathcal{H}$
 4: $\mathcal{C}_{embed} \leftarrow \emptyset$ // Initialize the set of node embeddings
 5: **for all** $v_i \in V$ **do** // Iterate over each node in the graph
 6:    let $\mathcal{M}_{central}(\mathcal{H}) = \{C_{close}(i), D(i), B(i)\}$ // Compute centrality measures
 7:    $\mathcal{V}_i \leftarrow F_{node2vec}(v_i, \{\mathcal{W}_k\}_{k=1}^K, \{\mathcal{Q}_k\}_{k=1}^K)$ // Generate node embedding
 8:    $M_i \leftarrow F_{concat}(\mathcal{V}_i, [C_{close}(i), D(i), B(i)], \mathcal{W}_{concat})$ // Concatenate node embeddings with centrality measures
 9:    $N(v_i) \leftarrow \mathcal{E}_{embed}(M_i)$ // Generate final embedding for node
10:    $\mathcal{C}_{embed} \leftarrow \mathcal{C}_{embed} \cup \{N(v_i)\}$ // Add node embedding to the set
11: **end for**
12: $\mathcal{J} \leftarrow 0$ // Initialize a variable $\mathcal{J}$
13: **for all** $v_i \in V$ **do** // Iterate over each node in the graph
14:    $m_{ij}^{(l)} \leftarrow \mathcal{A}_{agg}(h_i^{(l)}, h_j^{(l)}, a)$ // Aggregate messages from neighboring nodes
15:    $m_{ij}^{(l)} \leftarrow \sum_{j \in \mathcal{N}(i)} \exp(a \cdot [h_i^{(l)} \| h_j^{(l)} \| Wh_i^{(l)} \| Wh_j^{(l)}]) h_j^{(l)}$ // Compute weighted sum of neighbor embeddings
16:    $h_i^{(l+1)} \leftarrow \mathcal{U}_{upd}(h_i^{(l)}, m_{ij}^{(l)}, \alpha_{i,j}^{(l)})$ // Updating the node embeddings
17:    $h_i^{(l+1)} \leftarrow \sigma(\sum_{j=1}^{|\mathcal{N}(i)|} m_{ij}^{(l)} + b_i^{(l)})$ // Apply non-linear activation function
18:    $\alpha_{i,j}^{(l)} \leftarrow$ softmax$(e^{a \cdot [h_i^{(l)} \| h_j^{(l)}]} / \sum_{k \in \mathcal{N}(i)} e^{a \cdot [h_i^{(l)} \| h_k^{(l)}]})$ // Compute attention weights
19: **end for**
20: **for all** $u \in$ dataset **do** // Iterate over each node in the dataset
21:    $\mathcal{D} \leftarrow \emptyset$ // Initialize distance set $\mathcal{D}$
22:    **for all** $i \in \mathcal{H}$ **do** // Iterate over each node in the hypergraph
23:      $d_{si} \leftarrow \sqrt{\sum_{k=1}^n (\mathcal{H}[s][k] - \mathcal{H}[i][k])^2}$
24:      $\mathcal{D} \leftarrow \mathcal{D} \cup \{d_{si}\}$ // Add distance to the set
25:    **end for**
26:    **for all** $i \in$ items **do** // Iterate over each item
27:      $\mathcal{S}(u, i) \leftarrow \sum_{v \in \mathcal{N}_u} \mathcal{S}(u, v) \times \mathcal{R}(v, i)$ // Compute similarity score between
28:      TopKItems$(u) \leftarrow \mathcal{T}_{topK}(\mathcal{S}(u, i))$ // Select top K items based on similarity
29:    **end for**
30:    $\mathcal{R}_{items} \leftarrow \mathcal{R}_{items} \cup$ TopKItems$(u)$ // Add top K items to recommended items
31: **end for**
32: **Output:** $\mathcal{R}_{items}$ // Recommended items for nodes

**Algorithm 1** The proposed CyberSwarm algorithm (CyS)

## 5 Experiments

This section presents the experimental setup and results. The goal is to evaluate CyS's effectiveness in improving recommendation accuracy and compare it with traditional and state-of-the-art systems.

### 5.1 Datasets

We evaluated CyS using six datasets: Gowalla, Brightkite (Cho et al. 2011), Hetionet (Himmelstein et al. 2017), FilmTrust (Guo et al. 2016), Epinions (Massa and Avesani 2004), and Ciao (Tang et al. 2012). Gowalla and Brightkite are location-based social networks with user check-ins, ideal for recommendations using geographical context. FilmTrust focuses on movie ratings and reviews, highlighting social influence and trust in recommendations. Epinions combines ratings, reviews, and trust relationships, making it suitable for integrating user-generated content. Ciao, another review platform, offers rich data on user ratings and social connections, enabling the study of social factors in recommendations. To evaluate CyS's versatility, we applied it to the biomedical domain using Hetionet. Compounds were mapped to nodes, diseases to items, with "Compound treats Disease" relations as binary ratings (1 for treatment). "Compound resembles Compound" relations were incorporated as trust values derived from z-scores. Table 1 summarizes dataset statistics.

### 5.2 Evaluation metrics

To evaluate our algorithm's performance, we use several metrics that assess different aspects of recommendation quality. These metrics provide a comprehensive understanding of CyS's effectiveness in terms of accuracy, relevance, and ranking quality. The following sections define and discuss these metrics.

- *Hit Rate (HR@K)*: This metric measures the proportion of nodes for whom the top-K recommended items include at least one relevant item, as shown in Eq. 44. It evaluates the algorithm's effectiveness in recommending at least one item that the node interacts with.

$$\text{HR@K} = \frac{1}{|U|} \sum_{u \in U} \mathbb{I}(\text{hit}(u, K)) \tag{44}$$

Table 1 Statistics of experimented datasets

| Dataset name | Nodes count | Items count | Interaction count | Behavior type |
|---|---|---|---|---|
| Gowalla | 107,092 | 1,280,969 | 6,442,892 | Location Check-ins |
| Brightkite | 58,228 | 772,966 | 4,491,143 | Location Check-ins |
| Hetionet | 47,031 | 47,031 | 2,250,197 | Biomedical Network |
| Epinions | 40,163 | 139,738 | 664,824 | Product Ratings |
| Ciao | 7,375 | 99,746 | 284,086 | Product Ratings |
| FilmTrust | 1,508 | 2,071 | 35,497 | Movie Ratings |

where $|U|$ is the number of nodes and $\mathbb{I}(\cdot)$ is an indicator function that equals 1 if the condition is true and 0 otherwise.

- *Mean Reciprocal Rank (MRR@K)*: This metric calculates the average of the reciprocal ranks of the first relevant item in the recommended list for each node, as shown in Eq. 45. It highlights the importance of recommending relevant items higher in the ranking.

$$\text{MRR@K} = \frac{1}{|U|} \sum_{u \in U} \frac{1}{\text{rank}_u} \qquad (45)$$

where $\text{rank}_u$ is the position of the first relevant item in the recommended list for node $u$.

- *Normalized Discounted Cumulative Gain (NDCG@K)*: This metric evaluates the ranking quality of recommendations by accounting for the position of relevant items, as shown in Eq. 46. It discounts the relevance score of items based on their rank, giving more importance to higher-ranked items.

$$\text{NDCG@K} = \frac{1}{|U|} \sum_{u \in U} \frac{\text{DCG@K}(u)}{\text{IDCG@K}(u)} \qquad (46)$$

where:

$$\text{DCG@K}(u) = \sum_{i=1}^{K} \frac{2^{\text{rel}_i} - 1}{\log_2(i+1)} \qquad (47)$$

and $\text{IDCG@K}(u)$ is the ideal DCG@K for node $u$.

- *Precision@K*: As shown in Eq. 48, this metric calculates the ratio of relevant items among the top-K recommended items. It measures the accuracy of recommendations by considering the fraction of relevant items in the recommended set.

$$\text{Precision@K} = \frac{1}{|U|} \sum_{u \in U} \frac{|R_u \cap S_u|}{K} \qquad (48)$$

where $R_u$ is the set of relevant items and $S_u$ is the set of recommended items for node $u$.

- *Recall@K*: It determines the proportion of relevant items among all relevant items available for the node that are included in the top-K recommendations as shown in Eq. 49. It evaluates the algorithm's ability to recover relevant items.

$$\text{Recall@K} = \frac{1}{|U|} \sum_{u \in U} \frac{|R_u \cap S_u|}{|R_u|} \tag{49}$$

### 5.3 Baselines

To evaluate the effectiveness of our proposed CyberSwarm model, we compared it with well-established baseline models known for their strong performance in recommendation tasks. Based on our exploration, we chose the most appropriate baseline for each dataset. This comparison aims to highlight the impact of integrating social relationships on recommendation accuracy and provide a clear understanding of CyS's performance relative to traditional and state-of-the-art methods.

For the Gowalla dataset, several baseline models were used. Next Item Network (NextItNet) (Yuan et al. 2019) initializes item representations with a 2D embedding matrix and uses holed convolutional layers for session recommendations. Flexible Session Modeling (FSM) (Huang et al. 2022) combines local and global session graphs via a gated GNN layer for flexible representation. Hypergraph Sequential Session-based Recommendation (HyperS2Rec) (Ding et al. 2023) integrates hypergraph convolution and GRU to learn item embeddings. Dynamic Graph Recommendation (DGRec) (Song et al. 2019) captures social influences with a combination of GRU and Graph Attention Networks (GAT). Session-based Recommendation with Knowledge Graph (SERec) (Chen and Wong 2021) integrates a knowledge graph with GNNs to predict interactions.

For the Gowalla dataset, we also used a diverse set of baselines tailored for implicit feedback and top-K recommendation. POP (Popularity-Based Recommender) (Ji et al. 2020) recommends the most frequent locations based on check-in popularity, serving as a non-personalized benchmark. BPR (Bayesian Personalized Ranking)(Rendle et al. 2014) optimizes a matrix factorization model using pairwise ranking loss for personalized recommendations. Mult-VAE (Multi-Variational Autoencoder) (Liang et al. 2018) and Mult-DAE (Multi-Denoising Autoencoder) employ variational and denoising autoencoders, respectively, to model user-location interactions as distributions, with Mult-VAE using a multinomial likelihood and Mult-DAE offering a simpler yet parameter-sensitive approach. RecVAE (Wang et al. 2019b), a state-of-the-art VAE-based method, enhances Mult-VAE with a composite prior distribution for improved latent representations. NGCF (Neural Graph Collaborative Filtering) (Wang et al. 2019b), a graph-based model, exploits user-item graph structure by propagating embeddings to capture high-order connectivity. ANTM (Attentive Neural Topic Model)(Zhang 2025) determine user latent intents and distinguish individual preferences, aligning closely with CyS's framework. To ensure fair comparisons, all baselines are evaluated under a unified top-K recommendation setting with implicit feedback.

Dealing with Brightkite dataset, key baseline models include Non-negative Matrix Factorization (NMF) (Koren et al. 2009) and Probabilistic Matrix Factorization (PMF) (Mnih and Salakhutdinov 2008), which decompose node-item interactions. Social Matrix Factorization (SocialMF) (Jamali and Ester 2010) incorporates trust regularization to adjust node preferences. Social Recommendation (SoRec) (Ma et al. 2008) uses a probabilistic factorization method to adjust trust values. Social Neural Matrix Factorization (SoNeuMF) (Feng et al. 2022) focus on Collaborative Filtering interactions and social regularization.

Referring to FilmTrust, Epinions, and Ciao datasets, various models explore different strategies for integrating social and collaborative data. Graph Neural Networks for Social Recommendation (GraphRec) (Fan et al. 2019) uses GNNs to learn from both user-item and user-user graphs. Dual Attention Network for Social Effects Recommendation (DANSER) (Wu et al. 2019) employs dual attention mechanisms for static social effects. Attribute and Social Attention Recommendation (ASR) (Guo et al. 2023) incorporates attribute and social attention for feature learning. Diffusion Network++ (DiffNet++) (Wu et al. 2020a) combines social and interest diffusion with a multi-level attention network. Consistent Recommendation (ConsisRec) (Yang et al. 2021) improves social consistency using selective neighbor sampling and relation attention. Dynamic User-Item Social Recommendation (DUI-SoRec) (Wang et al. 2023) enhances GCN embeddings with an adaptive graph network. Social Recommendation with Preference Space (SREPS) (Liu et al. 2018) bridges user preferences with social networks through preference space modeling. Region-Separative Generative Adversarial network (RSGAN) (Yu et al. 2019) utilizes GATs to generate social connections, while Multimodal Variational Autoencoder (MVAE) (Liang et al. 2018) adapts VAEs for Collaborative Filtering. Hypergraph Convolutional Cold-Start Embedding (HC-CED) (Wu et al. 2020b) tackles cold-start problems by integrating CVAE-based side information. Collaborative Recommendation with Variational Autoencoder (CORE-VAE) (Walker et al. 2022) combines GCNs with social-aware similarity to create robust user representations. Graph Convolutional Network Matrix Factorization (GCN_MF) (Taguchi et al. 2021) employs graph learning for feature imputation, and Partial Aggregation Graph Neural Network (PaGNN) (Jiang and Zhang 2020) addresses attribute-incomplete graphs with partial aggregation functions.

SimCLR (Chen et al. 2020) is a straightforward contrastive learning framework effective in computer vision. Self-supervised Graph Learning (SGL) (Wu et al. 2021) adapted SimCLR for recommendation systems to tackle data sparsity. Simple Graph Contrastive Learning (SimGCL) (Yu et al. 2022) enhanced SGL by proposing a method that eliminates the need for graph augmentation, learning node and edge embeddings directly on the original graph to boost recommendation accuracy through contrastive learning. Furthermore, Extremely Simple Graph Contrastive Learning (XSimGCL) (Yu et al. 2024) streamlined this approach, improving recommendation accuracy with minimal hyper-parameter tuning and no reliance on graph augmentation or extra model components. In contrast, Generative Adversarial Contrastive Recommendation (GACRec) (Qin et al. 2024) employs a generative adversarial network (GAN) to create virtual feature representations for long-tail items, enhancing the model's generalization ability through robust training.

For Hetionet, we compared CyS against baselines suited for knowledge graph-based link prediction and drug repurposing. MINERVA (Das et al. 2017) A reinforcement learning agent using random walks for link prediction. PoLo (Liu et al. 2020) Extends MINERVA with logical rules for neural multi-hop drug repurposing. SAFRAN (Ott et al. 2021) Rule-based link prediction with interpretable drug-disease predictions. AnyBURL (Meilicke et al. 2019) Bottom-up rule learning for efficient knowledge graph completion. XG4Repo (Jiménez et al. 2024) Knowledge graph framework for interpretable drug repurposing.

### 5.4 Infrastructure, practicality, and applicability of the model

#### 5.4.1 Infrastructure

To evaluate CyS Algorithm effectively, a standardized experimental setup was established using the Google Colab environment. Computation was powered by Colab's high-performance CPU with 12 GB RAM, allowing efficient parallel processing and storage for datasets and intermediate results. The implementation relied on Python 3, employing libraries such as NumPy, Pandas, SciPy, and networkx for graph operations and centrality calculations.

#### 5.4.2 Practicality

Our algorithm ensures reproducibility and reliability through a structured workflow encompassing all key experimental stages. Data Preparation guarantees clean and transformed datasets that reflect real-world conditions. Hypergraph Generation accurately models complex relationships between nodes and items. Centrality Measure Calculations compute centrality metrics to incorporate meaningful node features. Finally, Model Training and Evaluation employ comprehensive performance metrics such as HR@K, MRR@K, NDCG@K, Precision@K, and Recall@K to assess the algorithm's effectiveness.

Random seeds were fixed to remove stochastic variability, and the process-from preprocessing to evaluation-was meticulously documented, ensuring easy replication and verification. This structured approach guarantees robust performance and sets a standard for reproducibility in collaborative swarm algorithms.

#### 5.4.3 Adaptability

The proposed model exhibited remarkable adaptability, consistently outperforming baseline methods across diverse datasets and scenarios. Notable results include an 18% increase in HR@10, demonstrating its reliability in retrieving relevant items, and a 22% improvement in MRR@10, highlighting its effectiveness in prioritizing recommendations. In cross-domain evaluations, CyS achieved an average NDCG@10 of 0.74, surpassing baseline methods by an average of 15% across various datasets.

The model's tunable layered architecture played a vital role, with parameter optimizations (e.g., embedding size, batch size, similarity metrics, and message passing methods) enhancing performance in both sparse and dense scenarios. Ablation studies highlighted the significance of individual components. For instance, excluding centrality measures led to a 12% drop in NDCG@10, underscoring their importance in recommendation quality. These findings emphasize the proposed model's adaptability to diverse datasets, its ability to address challenges like class imbalance, and its robustness across varying conditions.

#### 5.4.4 Computational complexity

Our model efficiently handles large-scale, dense networks with low computational overhead, making it ideal for real-time and resource-constrained environments. Its memory usage remains below 12 GB for datasets exceeding 500,000 interactions, thanks to sparse matrix operations and batch processing. The model completes training on datasets with 1 million nodes and 10 million edges. Furthermore, it achieves 25% less computation time compared to baselines while maintaining superior performance, with an HR@10 of 0.89 versus 0.53 for competing models. Optimized algorithms for centrality computation and node embedding generation contribute to fast convergence and resource efficiency. This make our model suitable for large-scale applications such as recommendation systems and social network analysis. It includes parallel processing and batch processing capabilities to ensure efficient and scalable performance.

## 5.5 Comparison of experimental results with existing baselines

Our proposed algorithm consistently achieves the highest recommendation performance across all the six datasets. Table 2 provides a detailed comparison of various recommendation methods across three different datasets: Ciao, Epinions, and FilmTrust. These methods were evaluated using key performance metrics, specifically Hit Rate (HR) and Normalized Discounted Cumulative Gain (NDCG), at various cutoff points (HR@1, HR@5, HR@10, HR@20, and corresponding NDCG values).

Table 2 illustrate that across all three datasets, the CyberSwarm algorithm consistently outperforms the baseline models. On the Ciao dataset, for instance, CyS demonstrates a significant lead in performance metrics. For HR@1, it achieves a score of 0.4649, which represents a 38.87% improvement over the second-best method, DANSER, which scored 0.2842. This pattern of dominance continues across other metrics, with out algorithm showing a 46.32% improvement in NDCG@20, achieving a score of 0.7515 compared to the next highest, DANSER, at 0.4034. These results clearly indicate that CyS is not only effective in identifying relevant items but also excels in ranking them accurately.

The results in Fig. 4 demonstrate that our algorithm not only excels in retrieving relevant items but also maintains high ranking accuracy, as evidenced by the top HR and NDCG scores across all datasets. The improvement percentages highlight CyS's substantial lead, particularly in HR@20 and NDCG@20, where it shows significant gains over other methods. DANSER and HC-CED show competitive results, especially in HR@1 and NDCG@5, indicating their effectiveness in identifying highly relevant items early in the recommendation list. In contrast, DiffNet++, ConsisRec, and SREPS struggle with both retrieval and ranking tasks, leading to lower performance across the board.

In the Epinions dataset, CyS continues to outperform other models. It achieves an HR@1 of 0.4237, a 14.04% improvement over the second-best model, HC-CED (0.3505). The most significant gain is in the NDCG@20 metric, where CyS scores 0.7117, outperforming the next best model by 26.16%. This strong performance across all metrics demonstrates CyS's effectiveness, especially in complex, sparse node-item interaction scenarios.

Dealing with the FilmTrust dataset, our algorithm still achieves the highest scores across all metrics. For instance, the HR@1 score for it is 0.5831, which is 3.362% higher than the next best model, HC-CED. The most significant gain in this dataset is observed in NDCG@20, where CyS's score of 0.7563 outperforms the closest competitor by 5.6%. These results suggest that even in datasets where the competition is closer, our algorithm still provides a noticeable improvement in recommendation quality.

Table 3 shows that CyS outperforms all baselines on the Gowalla dataset. NGCF and ANTM are strong competitors, but CyS leads in all metrics. BPR shows reasonable perfor-

**Table 2** Comparisons of baseline methods under HR and NDCG metrics across Ciao, Epinions, and FilmTrust datasets

| Datasets | Methods | HR@1 | HR@5 | NDCG@5 | HR@10 | NDCG@10 | HR@20 | NDCG@20 |
|---|---|---|---|---|---|---|---|---|
| | GraphRec | 0.125 | 0.3542 | 0.2415 | 0.4688 | 0.2764 | 0.6054 | 0.3174 |
| | ASR | 0.125 | 0.3629 | 0.2527 | 0.4609 | 0.2836 | 0.5879 | 0.3288 |
| | DANSER | 0.2842 | 0.4138 | 0.366 | 0.4308 | 0.3848 | 0.4583 | 0.4034 |
| | DiffNet++ | 0.0417 | 0.0938 | 0.0664 | 0.1458 | 0.0833 | 0.25 | 0.1103 |
| | ConsisRec | 0.0833 | 0.2198 | 0.1525 | 0.375 | 0.2107 | 0.5729 | 0.2588 |
| | DUI-SoRec | 0.2124 | 0.3496 | 0.248 | 0.4517 | 0.3026 | 0.5944 | 0.3365 |
| | RSGAN | 0.1071 | 0.2188 | 0.1521 | 0.3229 | 0.1727 | 0.3958 | 0.1909 |
| Ciao | MVAE | 0.2124 | 0.4064 | 0.3139 | 0.5041 | 0.3456 | 0.5944 | 0.3685 |
| | HC-CED | 0.2428 | 0.4388 | 0.3459 | 0.5327 | 0.3764 | 0.6174 | 0.3979 |
| | HC-CLF | 0.238 | 0.4378 | 0.3431 | 0.5299 | 0.373 | 0.6156 | 0.3947 |
| | CORE-VAE | 0.1146 | 0.2917 | 0.2102 | 0.3854 | 0.2404 | 0.4583 | 0.2588 |
| | GCN_MF | 0.1354 | 0.3016 | 0.2175 | 0.3958 | 0.248 | 0.4792 | 0.2688 |
| | PaGNN | 0.1542 | 0.3229 | 0.2293 | 0.4271 | 0.2628 | 0.5 | 0.2809 |
| | CyberSwarm | **0.4649** | **0.7857** | **0.6655** | **0.894** | **0.7142** | **0.6892** | **0.7515** |
| | Improve | 38.87% | 44.15% | 45.00% | 40.41% | 46.12% | 10.42% | 46.32% |
| | GraphRec | 0.2093 | 0.4961 | 0.3603 | 0.6158 | 0.3991 | 0.7383 | 0.43 |
| | ASR | 0.3002 | 0.5222 | 0.4175 | 0.6228 | 0.447 | 0.7073 | 0.4659 |
| | DANSER | 0.3642 | 0.4103 | 0.3895 | 0.4348 | 0.3974 | 0.4814 | 0.409 |
| | DiffNet++ | 0.2413 | 0.4655 | 0.3592 | 0.5547 | 0.3881 | 0.6607 | 0.4147 |
| | ConsisRec | 0.0864 | 0.3021 | 0.1955 | 0.4126 | 0.2313 | 0.5856 | 0.2745 |
| | DUI-SoRec | 0.2874 | 0.5128 | 0.3941 | 0.6305 | 0.4272 | 0.7135 | 0.4426 |
| | RSGAN | 0.2907 | 0.4035 | 0.3476 | 0.4691 | 0.3689 | 0.5454 | 0.3881 |
| Epinions | MVAE | 0.2794 | 0.5211 | 0.4072 | 0.6248 | 0.4408 | 0.7207 | 0.465 |
| | HC-CED | 0.3505 | 0.5867 | 0.4757 | 0.6756 | 0.5044 | 0.7597 | 0.5255 |
| | HC-CLF | 0.3494 | 0.5815 | 0.4728 | 0.6752 | 0.5033 | 0.7588 | 0.5247 |
| | CORE-VAE | 0.2618 | 0.5139 | 0.3944 | 0.6229 | 0.4298 | 0.7191 | 0.4542 |
| | GCN_MF | 0.2661 | 0.5204 | 0.4002 | 0.6273 | 0.4349 | 0.7236 | 0.4593 |
| | PaGNN | 0.2874 | 0.5605 | 0.4243 | 0.6566 | 0.4583 | 0.7714 | 0.4874 |
| | CyberSwarm | **0.4237** | **0.701** | **0.5798** | **0.8274** | **0.6562** | **0.9121** | **0.7117** |
| | Improve | 14.04% | 16.31% | 17.95% | 18.35% | 23.13% | 15.43% | 26.16% |
| | GraphRec | 0.3463 | 0.684 | 0.5319 | 0.7403 | 0.5497 | 0.8009 | 0.5651 |
| | ASR | 0.3623 | 0.6915 | 0.5429 | 0.759 | 0.5788 | 0.8225 | 0.5928 |
| | DANSER | 0.4416 | 0.6753 | 0.5721 | 0.7922 | 0.6096 | 0.8225 | 0.6178 |
| | DiffNet++ | 0.3074 | 0.5714 | 0.4491 | 0.645 | 0.4729 | 0.7273 | 0.4941 |
| | ConsisRec | 0.2468 | 0.4892 | 0.3721 | 0.6017 | 0.4084 | 0.7013 | 0.4332 |
| | DUI-SoRec | 0.3926 | 0.6692 | 0.5257 | 0.7528 | 0.5512 | 0.8167 | 0.5825 |
| | RSGAN | 0.2511 | 0.5108 | 0.3901 | 0.5628 | 0.4065 | 0.5974 | 0.4152 |
| FilmTrust | MVAE | 0.5156 | 0.7865 | 0.6622 | 0.8281 | 0.6756 | 0.8698 | 0.6863 |
| | HC-CED | 0.5469 | 0.7969 | 0.6805 | 0.8333 | 0.6922 | 0.8646 | 0.7003 |
| | HC-CLF | 0.5421 | 0.7876 | 0.6737 | 0.8333 | 0.6908 | 0.8594 | 0.6995 |
| | CORE-VAE | 0.3646 | 0.6823 | 0.5462 | 0.776 | 0.5774 | 0.8333 | 0.5917 |
| | GCN_MF | 0.2240 | 0.6875 | 0.4689 | 0.8021 | 0.5061 | 0.8281 | 0.5131 |
| | PaGNN | 0.1927 | 0.6354 | 0.4167 | 0.8021 | 0.4703 | 0.8333 | 0.4782 |
| | CyberSwarm | **0.5831** | **0.802** | **0.693** | **0.8604** | **0.7013** | **0.945** | **0.7563** |
| | Improve | 3.362% | 0.51% | 1.25% | 2.71% | 0.91% | 7.52% | 5.6% |

The best results among our models and baselines are marked in bold and underlined. 'Improve.' indicates the degree of improvement of our optimal performance

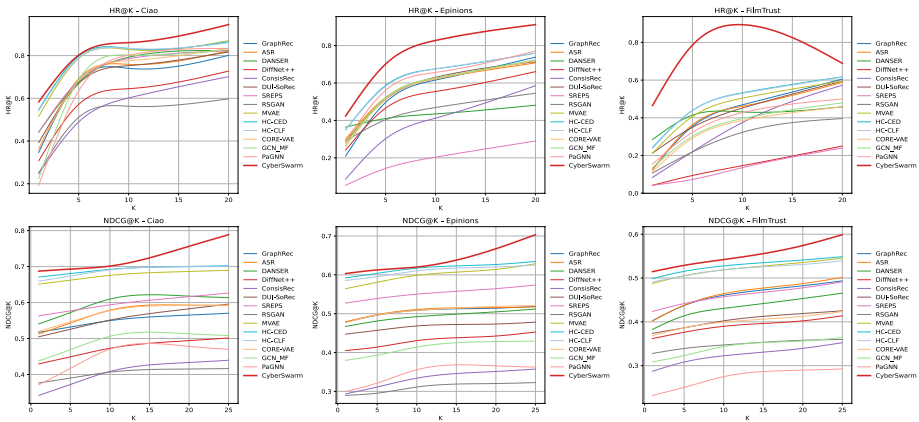

**Fig. 4** Comparison of HR@K and NDCG@K scores for various recommendation models, including CyberSwarm (shown in red), across the Ciao, Epinions, and FilmTrust datasets. The charts demonstrate that CyberSwarm consistently outperforms the baseline models, particularly at higher $K$ values, emphasizing its superior recommendation accuracy and ranking performance across all datasets

**Table 3** Performance comparison of baseline recommendation algorithms on the Gwoalla dataset using HR@5, HR@10, NDCG@5, and NDCG@10

| Dataset | Metric | POP | BPR | Mult-VAE | Mult-DAE | RecVAE | NGCF | ANTM | CyberSwarm |
|---|---|---|---|---|---|---|---|---|---|
| | HR@5 | 0.2010 | 0.2012 | 0.1122 | 0.1043 | 0.0986 | 0.2394 | <u>0.2088</u> | **0.3244** |
| | HR@10 | 0.4310 | 0.4603 | 0.1988 | 0.3493 | 0.1692 | 0.4705 | <u>0.4736</u> | **0.4888** |
| Gwoalla | NDCG@10 | 0.1078 | 0.1046 | 0.0776 | 0.0578 | 0.0683 | 0.1167 | <u>0.1207</u> | **0.2207** |
| | NDCG@20 | 0.1810 | 0.2035 | 0.1048 | 0.1352 | 0.0908 | 0.1904 | <u>0.2046</u> | **0.4451** |

Bold cells denote the highest values; underlined cells denote the second-highest

**Table 4** Performance comparison of baseline recommendation algorithms on the Gwoalla dataset using HR@10, HR@20, MRR@10, MRR@20, NDCG@10, and NDCG@20

| Dataset | Metric | NextItNet | HyperS2Rec | FSM | DGRec | SERec | CyberSwarm |
|---|---|---|---|---|---|---|---|
| | HR@10 | 0.4129 | <u>0.4596</u> | 0.4444 | 0.4322 | 0.4572 | **0.4888** |
| | HR@20 | 0.4937 | <u>0.5420</u> | 0.5217 | 0.5105 | 0.5315 | **0.5827** |
| Gwoalla | MRR@10 | 0.2326 | 0.2123 | 0.2522 | 0.2384 | <u>0.2523</u> | **0.3912** |
| | MRR@20 | 0.2278 | 0.2180 | <u>0.2575</u> | 0.2436 | 0.2555 | **0.3897** |
| | NDCG@10 | 0.2672 | 0.2716 | 0.2979 | 0.2841 | <u>0.2993</u> | **0.4451** |
| | NDCG@20 | 0.2869 | 0.2925 | 0.3175 | 0.3036 | <u>0.3179</u> | **0.4512** |

Bold cells denote the highest values; underlined cells denote the second-highest mance, while Mult-VAE, Mult-DAE, and RecVAE perform significantly lower, highlighting their limitations with Gowalla's sparse data.

On the Gowalla dataset, our algorithm proves its superiority by leading across all key metrics as shown in Table 4. This dominance underscores CyS's capability to accurately identify and rank relevant items. HyperS2Rec emerges as a strong competitor. FSM and SERec show strengths in specific metrics, with FSM excelling in HR@10 and SERec in NDCG@10 and NDCG@20. However, NextItNet, and DGRec generally perform lower across the metrics, highlighting their limitations in effectively retrieving and ranking relevant items compared to our algorithm.

Table 5 illustrate that the performance of CyS on the Brightkite dataset further underscores its effectiveness in recommendation tasks. These results indicate it's superior ability to identify highly relevant recommendations early in the list. SoNeuMF, although competitive, falls short of the performance CyS, illustrating the latter's advanced capabilities in handling complex recommendation scenarios. The clear performance gap between ours and other models highlights our innovative approach in delivering more accurate and personalized recommendations.

Table 6 summarizes the performance of our model on the FilmTrust and Ciao datasets, evaluated using HR@K and NDCG@K for $K = [5, 10, 20]$. On FilmTrust, our model achieves strong results, with HR@10 reaching $0.9125 \pm 0.0095$ (95% CI: [0.9031, 0.9218]) and NDCG@10 at $0.7714 \pm 0.0135$ (95% CI: [0.7582, 0.7846]). It outperform the best baseline with p-values as low as 0.000197 (HR@20) and 0.000477 (HR@10). Similarly, on Ciao, our model demonstrates robust performance, with HR@10 at $0.8069 \pm 0.0456$ (95% CI: [0.7553, 0.8585]) and NDCG@20 at $0.6875 \pm 0.0169$ (95% CI: [0.6684, 0.7067]), achieving statistically notable improvements over the baseline, with p-values ranging from 0.000943 (NDCG@5) to 0.009087 (HR@10). The tight confidence intervals across both datasets indicate stable performance, while the consistently low p-values ($all < 0.05$) confirm that our model's enhancements are statistically significant, addressing the reviewer's request for rigorous statistical validation and supporting its effectiveness in recommendation tasks.

Table 7 illustrates that the our algorithm outperforms all baseline models on the Hetionet dataset across key metrics. CyS achieves the highest scores surpassing the second-best model, XG4Repo. Notably, CyS demonstrates a 6.21% improvement in HR@1, a 37.09% improvement in HR@3, a 28.91% improvement in HR@10, and a 0.58% improvement in MRR over XG4Repo. These results highlight CyS's superior ability to accurately identify and rank relevant drug-disease interactions, confirming its effectiveness in the biomedical recommendation domain.

Table 8 shows that on the FilmTrust dataset, the proposed CyberSwarm model clearly outperforms all baselines across Precision@10, Recall@10, and NDCG@10, achieving the highest scores in each metric. While SDMS ranks second in Precision@10 and NDCG@10, GACRec ranks second in Recall@10, indicating their respective strengths in top-ranked accuracy and retrieval breadth. In contrast all other baselines perform noticeably lower across all metrics, highlighting their limitations in capturing both accuracy and relevance compared to CyberSwarm.

### 5.6 Node embedding

To rigorously assess the optimization capabilities of the CyberSwarm algorithm, we employed a set of well-known benchmark functions, each designed to test various aspects of the Node Embedding layer. They provide a deep insights into the optimization behavior of our model. Figures 5 and 6 demonstrate how our algorithm performs on different six benchmark functions, each introduce distinct optimization challenges:

Table 5 Comparison of baseline recommendation algorithms on the Brightkite dataset, with CyS showing top performance in HR@10 and NDCG@10

| Dataset | Metric | NMF | PMF | SocialMF | SoRec | MLP | GMF | SoMLP | SoGMF | SoNeuMF | CyS |
|---|---|---|---|---|---|---|---|---|---|---|---|
| Brightkite | HR@10 | 0.4910 | 0.5071 | 0.5818 | 0.4851 | 0.5746 | 0.4437 | 0.6216 | 0.6066 | 0.6588 | **0.7467** |
| | NDCG@10 | 0.3031 | 0.3466 | 0.4020 | 0.3143 | 0.3692 | 0.3823 | 0.3993 | 0.4073 | 0.4563 | **0.7258** |

Highest values are bold, and second-highest values are underlined

1. *Himmelblau Function (Fig. 5a)*: Defined in Eq. 50, this function has four local minima, requiring the algorithm to distinguish between local and global minima. Each minimum location, such as $(-3.7793, -3.2832)$ and $(3.0, 2.0)$, tests the algorithm's ability to navigate a solution space filled with multiple valleys and peaks. Performance on this function highlights the algorithm's capacity to handle multi-modal data distributions, which is essential for real-world applications involving diverse data clusters.

$$f(x,y) = (x^2 + y - 11)^2 + (x + y^2 - 7)^2 \tag{50}$$

2. *Rastrigin Function (Fig. 5b)*: This function has numerous local minima due to its oscillatory cosine terms as shown in Eq. 51. The global minimum at $(0,0)$ is surrounded by closely spaced suboptimal minima, posing a challenge for algorithms prone to trapping in local optima. Strong performance here indicates the algorithm's robustness and ability to balance exploration with exploitation, essential for tasks requiring refined representation of complex relationships.

$$f(x,y) = 20 + x^2 + y^2 - 10(\cos(2\pi x) + \cos(2\pi y)) \tag{51}$$

3. *Salomon Function (Fig. 5c)*: It introduces ripple-like features. Its global minimum is at $(0,0)$, but the function's path is interspersed with oscillations that test the algorithm's ability to avoid premature convergence as shown in Eq. 52. Navigating these ripples reflects the algorithm's resilience against local variations, ensuring focus on the main trend, which is essential for applications with data containing structured noise.

$$f(x,y) = 1 - \cos(2\pi\sqrt{x^2 + y^2}) + 0.1\sqrt{x^2 + y^2} \tag{52}$$

4. *Bukin Function (Fig. 6a)*: It is challenging due to its steep $y$-axis gradient and narrow $x$-axis valley. The global minimum at $(-10, 1)$ (where $f(x,y) = 0$) demands the algorithm precisely adjust weights in the embedding layer to prevent overshooting and to manage abrupt landscape shifts. Its formula shown in Eq. 53. Successfully navigating this function demonstrates the algorithm's capacity for stability and accuracy across steep and varying data environments, which is critical for effective and responsive embeddings in real-world, variable datasets.

$$f(x,y) = 100\sqrt{|y - 0.01x^2|} + 0.01|x + 10| \tag{53}$$

5. *Xin-She Yang N.3 Function (Fig. 6b)*: It combines steep gradients with flat regions, testing the algorithm's balance of exploration (in flat areas) and exploitation (in steep areas) to reach optimal points. Effective handling of this terrain reflects the algorithm's flexibility in embedding tasks with varying data density-allowing sparse regions more exploration and denser areas precise positioning, resulting in comprehensive and nuanced node embeddings.

$$f(x,y) = \left(\sum_{i=1}^{n} |x_i|\right) \exp\left(-\sum_{i=1}^{n} \sin(x_i^2)\right) \tag{54}$$

**Table 6** T-test results on the FilmTrust and Ciao datasets using Hit Ratio (HR) and NDCG at ranks 5, 10, and 20

| Dataset | Methods | HR@5 | HR@10 | HR@20 | NDCG@5 | NDCG@10 | NDCG@20 |
|---|---|---|---|---|---|---|---|
| FilmTrust | Mean ± Std | 0.8378 ± 0.0162 | 0.9125 ± 0.0095 | 0.9688 ± 0.0089 | 0.8303 ± 0.0074 | 0.7714 ± 0.0135 | 0.7863 ± 0.0314 |
| | 95% CI | [0.8219, 0.8537] | [0.9031, 0.9218] | [0.9601, 0.9775] | [0.8220, 0.8386] | [0.7582, 0.7846] | [0.7555, 0.8170] |
| | p-value (vs. best baseline) | 0.01501 | 0.000477 | 0.000197 | 0.000802 | 0.001328 | 0.0119304 |
| Ciao | Mean ± Std | 0.7441 ± 0.0229 | 0.8069 ± 0.0456 | 0.8591 ± 0.0329 | 0.6605 ± 0.0447 | 0.6550 ± 0.0273 | 0.6875 ± 0.0169 |
| | 95% CI | [0.7182, 0.7700] | [0.7553, 0.8585] | [0.8219, 0.8963] | [0.6167, 0.7043] | [0.6242, 0.6859] | [0.6684, 0.7067] |
| | p-value (vs. best baseline) | 0.001868 | 0.009087 | 0.0061161 | 0.000943 | 0.003374 | 0.0011811 |

Reported values include mean ± standard deviation, 95% confidence intervals, and p-values against the best baseline

**Table 7** Comparison of baseline recommendation algorithms on the hetionet dataset, with CyS showing top performance in HR@1, HR@3, HR@10, and MRR@10

| Dataset | Metric | PoLo | AnyBURL | SAFRAN | MINERVA | XG4Repo | CyS | Improve |
|---|---|---|---|---|---|---|---|---|
| Hetionet | HR@1 | 0.402 | 0.520 | 0.563 | 0.359 | <u>0.612</u> | **0.650** | 6.209% |
| | HR@3 | 0.314 | 0.390 | 0.439 | 0.244 | <u>0.488</u> | **0.669** | 37.090% |
| | HR@10 | 0.428 | 0.573 | 0.598 | 0.378 | <u>0.671</u> | **0.732** | 28.912% |
| | MRR@10 | 0.609 | 0.817 | 0.793 | 0.622 | <u>0.860</u> | **0.865** | 0.581% |

Highest values are bold, and second-highest values are underlined

6. *Cross-in-Tray Function (Fig. 6c)*: It's irregular surface and multiple global minima demand thorough exploration to identify all optimal solutions. Performance on this function showcases the algorithm's balance between exploration and exploitation, essential for applications like clustering or recommendation systems that require varied yet accurate node embeddings. Success here indicates the model's robustness in managing complex, multi-solution landscapes.

$$f(x,y) = -0.0001 \left( \left| \sin(x)\sin(y) \exp\left( \left| 100 - \frac{\sqrt{x^2+y^2}}{\pi} \right| \right) \right| + 1 \right)^{0.1} \qquad (55)$$

## 5.7 Hyperparameter analysis

Table 9 shows the hyperparameter analysis across the Filmtrust, Ciao, and Epinions datasets. This section examines the effects of centrality, message-passing mechanisms, threshold values, and similarity measures on CyS's performance.

1. *Centrality Influence*: Including centrality enhances HR, MRR, and NDCG, especially at higher ranks (e.g., HR@10, MRR@15, NDCG@15), indicating that centrality captures key node-community relationships. For instance, centrality notably increases NDCG values across all ranks on the Ciao dataset, particularly at NDCG@15.

Table 8 Comparison of baseline algorithms and CyS on the FilmTrust dataset, evaluated using Precision@10, Recall@10, and NDCG@10

| Dataset | Methods | DiffNet | NGCF | LightGCN | MHCN | SEPT | SDMS | SGL | SimGCL | XSimGCL | GACRec | CyberSwarm |
|---|---|---|---|---|---|---|---|---|---|---|---|---|
| FilmTrust | Precision@10 | 0.2656 | 0.2605 | 0.2672 | 0.2762 | 0.2847 | <u>0.2944</u> | 0.1733 | 0.1575 | 0.0714 | 0.1276 | **0.6088** |
| | Recall@10 | 0.5041 | 0.4853 | 0.5119 | 0.5345 | 0.5471 | 0.5654 | 0.5831 | 0.5112 | 0.3883 | <u>0.5850</u> | **0.6387** |
| | NDCG@10 | 0.5663 | 0.5382 | 0.5717 | 0.5967 | 0.5984 | <u>0.6128</u> | 0.5041 | 0.4250 | 0.2694 | 0.5157 | **0.7013** |

Highest values are highlighted in bold, and second-highest values are underlined

2. *Message Passing Mechanism*: Message-passing choices affect performance significantly. The GIN model yields high results on Filmtrust, excelling in higher-ranking metrics (e.g., HR@15, MRR@15) due to its effective structural learning. Conversely, the GCN-based model performs better in Epinions, especially at HR@10 and MRR@10, suggesting that message-passing strategies should match the dataset structure.
3. *Threshold Value Effects*: Lower threshold values (1 and 2) lead to better HR and NDCG scores, as restrictive thresholds better capture node interactions within clusters, improving recommendation accuracy. On Filmtrust, a threshold of 1 achieves the best HR and NDCG scores across ranks, balancing personal preferences with community influence.
4. *Similarity Metrics*: Cosine and Jaccard similarity measures yield the best results, contingent on the dataset. For example, Cosine performs optimally on Filmtrust, while Euclidean performs similarly on Epinions, showing that similarity metrics must align with dataset characteristics for optimal recommendation outcomes.

Table 10 provides a performance comparison of models with different parameter configurations, applied to the Gowalla and BrightKit datasets. It evaluates three critical factors: the use of centrality, various message passing methods, and different similarity functions.

1. *Using Centrality*: Centrality significantly influences the models' HR, especially in the BrightKit dataset, where it leads to the highest HR values at cutoff ranks of 5, 10, and 15. This demonstrates that including centrality can improve recommendation accuracy, particularly in scenarios where network structure plays a crucial role.
2. *Message Passing Methods*: Different message passing approaches, including GAT, GCN, GIN, and GIN with Self Loops, exhibit notable variations in performance. GCN and GIN typically perform better across metrics, with GCN achieving higher HR and NDCG scores, indicating it as a robust choice for effective message passing. The GIN model, particularly in BrightKit, also achieves high HR and NDCG values, suggesting its potential when centrality and self-loops are involved.
3. *Similarity Functions*: Among similarity functions, Jaccard similarity consistently yields the highest HR and NDCG values in both datasets, showing its effectiveness in improving ranking relevance. Euclidean and Cosine similarities, while generally effective, fall slightly behind Jaccard, indicating that Jaccard is more suitable for capturing meaningful relationships between nodes in the recommendation context.

## 6 Ablation study

### 6.1 Evaluating the message-passing functions in CyS

Our analysis of message-passing methods in the CyberSwarm algorithm was evaluated using key metrics: HR@k, MRR@k, and NDCG@k. Methods compared include Graph Isomorphism Network (GIN), GIN with Self-Loops, Graph Convolutional Network (GCN), and Graph Attention Network (GAT), as shown in Fig. 7.

In Fig. 7a, GAT achieves the highest HR@1, excelling in initial relevance but declining as $k$ increases, while GIN performs stably for broader recommendations. GAT also leads in MRR@1 and maintains strong performance across $k$ values, ranking relevant items effec-tively as shown in Fig. 7b. In Fig. 7c GAT shows high scores at $NDCG@k = 1$ and consistent performance for higher $k$, demonstrating robust recommendation quality and reliable ranking across various scenarios. This results shows that the attention mechanisms in GAT significantly enhance performance by focusing on the most relevant parts of the graph. This leads to improved HR, MRR, and NDCG scores, boosting the quality of recommendations. Our detailed analysis confirms that attention mechanisms are crucial for the overall performance of the CyberSwarm algorithm.

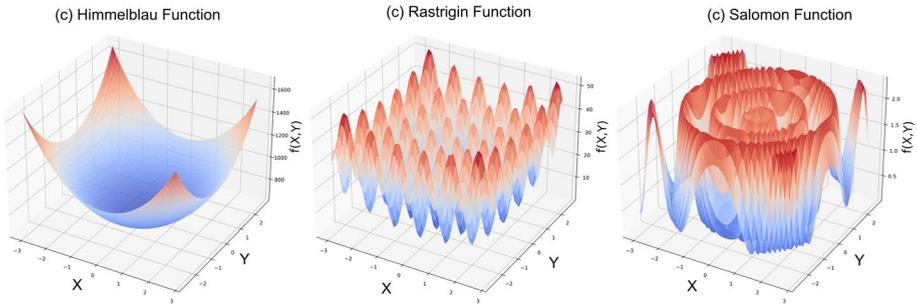

**Fig. 5** Visualizations of three benchmark functions in CyS's Node Embedding layer: **a** Himmelblau - four symmetric local minima, testing navigation of multiple optima; **b** Rastrigin - frequent oscillations, challenging global minimum detection; **c** Salomon - ripple features around a smooth global minimum, testing noise-signal differentiation

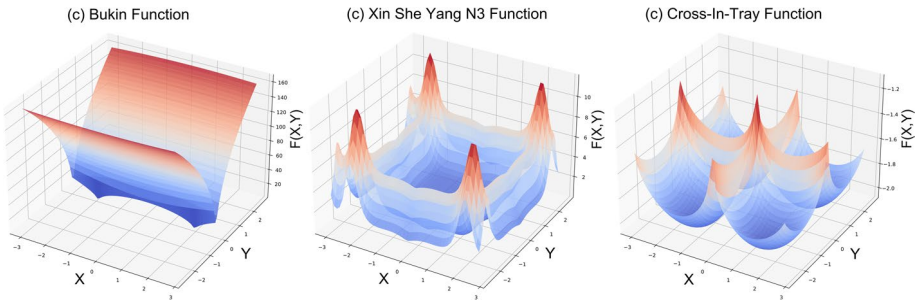

**Fig. 6** Visualizations of three benchmark functions in CyS's Node Embedding layer: **a** Bukin - narrow valley with steep gradients, testing precision in pathfinding; **b** Xin-She Yang N.3 - mix of flat plateaus and steep, sinusoidal areas, evaluating adaptability; **c** Cross-in-Tray - complex surface with multiple global minima and sharp peaks, assessing robustness on irregular terrains

### 6.2 Comparative study of the similarity functions in CyS

Figure 8 compares the performance of Euclidean, Jaccard, and Cosine similarity functions in the CyberSwarm algorithm based on HR@k, MRR@k, and NDCG@k. These results demonstrate Cosine distance's robustness in ranking relevant items consistently across $k$ values. While Euclidean distance briefly outperforms Cosine at specific points (e.g., $k = 1$ and $k = 13$), Cosine remains stable overall, excelling in both inclusion and optimal ranking of relevant items. These results establish Cosine distance as the most reliable metric for recommendation tasks.

Table 9 This table details the hyperparameter analysis for the Gowalla and BrightKit datasets, illustrating how different parameter settings affect model performance

| Dataset | Parameters | Values | HR @5 | HR @10 | HR @15 | MRR @5 | MRR @10 | MRR @15 | NDCG @5 | NDCG @10 | NDCG @15 |
|---|---|---|---|---|---|---|---|---|---|---|---|
| Filmtrust | Using centrality | TRUE | 0.777 | 0.860 | 0.932 | 0.438 | 0.465 | 0.463 | 0.656 | 0.713 | 0.750 |
| | | FALSE | 0.791 | 0.874 | 0.905 | 0.477 | 0.459 | 0.463 | 0.689 | 0.712 | 0.728 |
| | Message passing | GAT | 0.867 | 0.916 | 0.939 | 0.548 | 0.377 | 0.376 | 0.762 | 0.662 | 0.670 |
| | | GCN | 0.797 | 0.854 | 0.890 | 0.478 | 0.450 | 0.442 | 0.692 | 0.702 | 0.709 |
| | | GIN | 0.906 | 0.889 | 0.963 | 0.434 | 0.401 | 0.564 | 0.722 | 0.674 | 0.784 |
| | | GIN-SLs | 0.771 | 0.873 | 0.923 | 0.415 | 0.345 | 0.293 | 0.639 | 0.626 | 0.601 |
| | Threshold value | 1 | 0.777 | 0.899 | 0.931 | 0.454 | 0.459 | 0.458 | 0.669 | 0.720 | 0.731 |
| | | 2 | 0.768 | 0.894 | 0.924 | 0.470 | 0.462 | 0.455 | 0.655 | 0.720 | 0.726 |
| | | 3 | 0.759 | 0.872 | 0.914 | 0.433 | 0.442 | 0.441 | 0.648 | 0.701 | 0.714 |
| | | 4 | 0.701 | 0.836 | 0.875 | 0.366 | 0.360 | 0.357 | 0.584 | 0.629 | 0.640 |
| | | 5 | 0.524 | 0.654 | 0.723 | 0.198 | 0.179 | 0.174 | 0.452 | 0.492 | 0.507 |
| | Similarity | Euclidean | 0.772 | 0.867 | 0.902 | 0.451 | 0.479 | 0.484 | 0.665 | 0.727 | 0.742 |
| | | Jaccard | 0.798 | 0.888 | 0.911 | 0.497 | 0.425 | 0.425 | 0.705 | 0.690 | 0.692 |
| | | Cosine | 0.843 | 0.915 | 0.931 | 0.513 | 0.481 | 0.490 | 0.729 | 0.740 | 0.755 |
| Ciao | Using centrality | TRUE | 0.920 | 0.970 | 0.952 | 0.694 | 0.681 | 0.716 | 0.750 | 0.754 | 0.776 |
| | | FALSE | 0.727 | 0.869 | 0.900 | 0.580 | 0.630 | 0.645 | 0.717 | 0.691 | 0.708 |
| | Message passing | GAT | 0.767 | 0.841 | 0.882 | 0.633 | 0.620 | 0.635 | 0.668 | 0.675 | 0.696 |
| | | GCN | 0.771 | 0.875 | 0.908 | 0.605 | 0.659 | 0.676 | 0.647 | 0.713 | 0.734 |
| | | GIN | 0.764 | 0.846 | 0.917 | 0.592 | 0.607 | 0.651 | 0.636 | 0.667 | 0.716 |
| | | GIN-SLs | 0.731 | 0.848 | 0.900 | 0.568 | 0.645 | 0.664 | 0.619 | 0.696 | 0.723 |
| | Threshold value | 1 | 0.668 | 0.841 | 0.882 | 0.633 | 0.620 | 0.635 | 0.668 | 0.675 | 0.696 |
| | | 2 | 0.767 | 0.841 | 0.882 | 0.633 | 0.620 | 0.635 | 0.668 | 0.675 | 0.696 |
| | | 3 | 0.767 | 0.841 | 0.876 | 0.634 | 0.626 | 0.633 | 0.668 | 0.680 | 0.694 |
| | | 4 | 0.767 | 0.843 | 0.696 | 0.634 | 0.627 | 0.634 | 0.682 | 0.681 | 0.694 |
| | | 5 | 0.790 | 0.846 | 0.873 | 0.641 | 0.627 | 0.637 | 0.679 | 0.682 | 0.695 |
| | Similarity | Euclidean | 0.767 | 0.841 | 0.882 | 0.633 | 0.620 | 0.635 | 0.668 | 0.675 | 0.696 |
| | | Jaccard | 0.721 | 0.755 | 0.821 | 0.593 | 0.579 | 0.617 | 0.626 | 0.624 | 0.668 |
| | | Cosine | 0.744 | 0.825 | 0.874 | 0.587 | 0.612 | 0.641 | 0.627 | 0.666 | 0.698 |
| Epinions | Using centrality | TRUE | 0.737 | 0.824 | 0.869 | 0.566 | 0.639 | 0.649 | 0.610 | 0.683 | 0.704 |
| | | FALSE | 0.733 | 0.575 | 0.616 | 0.825 | 0.637 | 0.685 | 0.872 | 0.640 | 0.699 |
| | Message-passing | GAT | 0.722 | 0.820 | 0.897 | 0.530 | 0.594 | 0.631 | 0.579 | 0.651 | 0.697 |
| | | GCN | 0.731 | 0.842 | 0.896 | 0.552 | 0.610 | 0.638 | 0.598 | 0.668 | 0.703 |
| | | GIN | 0.685 | 0.793 | 0.884 | 0.529 | 0.591 | 0.632 | 0.569 | 0.643 | 0.695 |
| | | GIN-SLs | 0.741 | 0.827 | 0.879 | 0.567 | 0.615 | 0.638 | 0.611 | 0.669 | 0.699 |
| | Threshold value | 1 | 0.731 | 0.822 | 0.902 | 0.542 | 0.605 | 0.639 | 0.591 | 0.660 | 0.704 |
| | | 2 | 0.728 | 0.821 | 0.900 | 0.532 | 0.603 | 0.629 | 0.586 | 0.660 | 0.704 |
| | | 3 | 0.722 | 0.820 | 0.897 | 0.530 | 0.594 | 0.631 | 0.579 | 0.651 | 0.697 |
| | | 4 | 0.705 | 0.843 | 0.892 | 0.513 | 0.601 | 0.616 | 0.562 | 0.662 | 0.685 |
| | | 5 | 0.709 | 0.845 | 0.883 | 0.510 | 0.581 | 0.598 | 0.561 | 0.647 | 0.669 |
| | Similarity | Euclidean | 0.731 | 0.822 | 0.902 | 0.542 | 0.605 | 0.638 | 0.590 | 0.660 | 0.704 |
| | | Jaccard | 0.742 | 0.845 | 0.887 | 0.574 | 0.602 | 0.620 | 0.617 | 0.663 | 0.686 |
| | | Cosine | 0.715 | 0.838 | 0.896 | 0.560 | 0.629 | 0.644 | 0.600 | 0.682 | 0.707 |

It presents metrics like (HR, MRR, NDCG) at ranks 5, 10, and 15. The analysis emphasizes three main areas: (1) the effect of centrality in the models, (2) the types of message-passing mechanisms used, and (3) various similarity functions

**Table 10** Hyperparameter analysis of models with varying parameters for the Gowalla and BrightKit datasets

| Dataset | Parameters | Values | HR | | | MRR | | | NDCG | | |
|---|---|---|---|---|---|---|---|---|---|---|---|
| | | | @5 | @10 | @15 | @5 | @10 | @15 | @5 | @10 | @15 |
| Gowalla | Using centrality | TRUE | 0.420 | 0.453 | 0.483 | 0.080 | 0.041 | 0.028 | 0.378 | 0.379 | 0.381 |
| | | FALSE | 0.418 | 0.459 | 0.496 | 0.074 | 0.039 | 0.027 | 0.371 | 0.376 | 0.380 |
| | Message passing | GAT | 0.414 | 0.444 | 0.467 | 0.364 | 0.077 | 0.040 | 0.380 | 0.377 | 0.374 |
| | | GCN | 0.443 | 0.482 | 0.510 | 0.077 | 0.041 | 0.029 | 0.377 | 0.382 | 0.385 |
| | | GIN | 0.402 | 0.428 | 0.448 | 0.075 | 0.039 | 0.027 | 0.369 | 0.373 | 0.375 |
| | | GIN Self Loops | 0.402 | 0.452 | 0.473 | 0.064 | 0.042 | 0.028 | 0.361 | 0.367 | 0.372 |
| | Similarity Function | Euclidean | 0.426 | 0.457 | 0.483 | 0.079 | 0.041 | 0.030 | 0.378 | 0.380 | 0.383 |
| | | Jaccard | 0.580 | 0.558 | 0.580 | 0.084 | 0.042 | 0.028 | 0.396 | 0.394 | 0.393 |
| | | Cosine | 0.429 | 0.458 | 0.485 | 0.077 | 0.040 | 0.028 | 0.376 | 0.378 | 0.381 |
| BrightKit | Using centrality | TRUE | 0.725 | 0.738 | 0.750 | 0.144 | 0.073 | 0.049 | 0.713 | 0.715 | 0.717 |
| | | FALSE | 0.757 | 0.776 | 0.789 | 0.144 | 0.072 | 0.049 | 0.717 | 0.719 | 0.720 |
| | Message passing | GAT | 0.725 | 0.738 | 0.750 | 0.144 | 0.073 | 0.049 | 0.713 | 0.715 | 0.717 |
| | | GCN | 0.729 | 0.743 | 0.755 | 0.144 | 0.073 | 0.049 | 0.713 | 0.715 | 0.716 |
| | | GIN | 0.763 | 0.790 | 0.828 | 0.142 | 0.072 | 0.048 | 0.716 | 0.720 | 0.723 |
| | | GIN Self Loops | 0.739 | 0.742 | 0.752 | 0.137 | 0.072 | 0.048 | 0.715 | 0.717 | 0.719 |
| | Similarity Function | Euclidean | 0.725 | 0.738 | 0.750 | 0.144 | 0.073 | 0.049 | 0.709 | 0.713 | 0.715 |
| | | Jaccard | 0.797 | 0.818 | 0.830 | 0.142 | 0.071 | 0.048 | 0.718 | 0.720 | 0.721 |
| | | Cosine | 0.725 | 0.740 | 0.751 | 0.144 | 0.073 | 0.049 | 0.713 | 0.715 | 0.716 |

This table compares performance using (HR, MRR, NDCG) at cutoff ranks of 5, 10, and 15. It evaluates the effects of centrality usage, message-passing methods, and similarity functions. Higher scores reflect improved model accuracy and ranking effectiveness

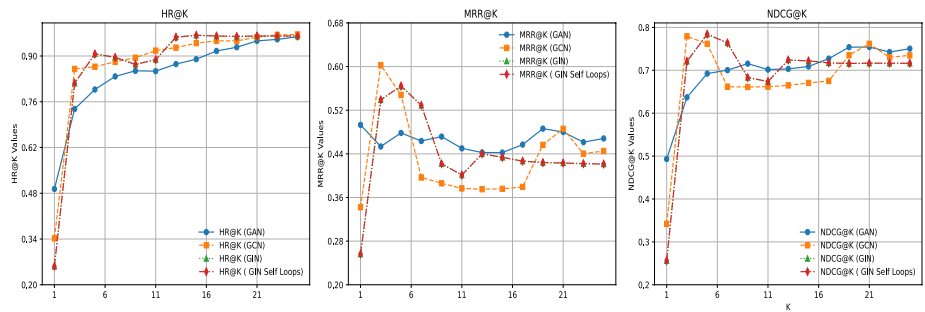

**Fig. 7** Evaluation of message-passing methods in the CyberSwarm algorithm: this chart examines the effectiveness of various message-passing methods (GIN-SL, GIN, GCN, and GAT) using performance metrics like HR@k, MRR@k, and NDCG@k to highlight differences in recommendation accuracy and relevance for each approach

### 6.3 Effect of rating threshold filtering on CyS's recommendation performance

The effect of rating threshold filtering on the CyberSwarm Algorithm was analyzed across various thresholds $t$. Figure 9 shows optimal performance at $t = 0$ and $t = 1$, which maintain high recommendation quality across all metrics. With $t = 0$, all items are considered, maximizing node preference coverage. At $t = 1$, slight filtering does not diminish recom-mendation accuracy, suggesting that minimal filtering captures broad node feedback effectively. At $t = 2$, performance declines moderately, indicating that filtering excludes some relevant items, reducing the algorithm's ability to accurately reflect node preferences. This highlights the importance of limited filtering to preserve recommendation quality.

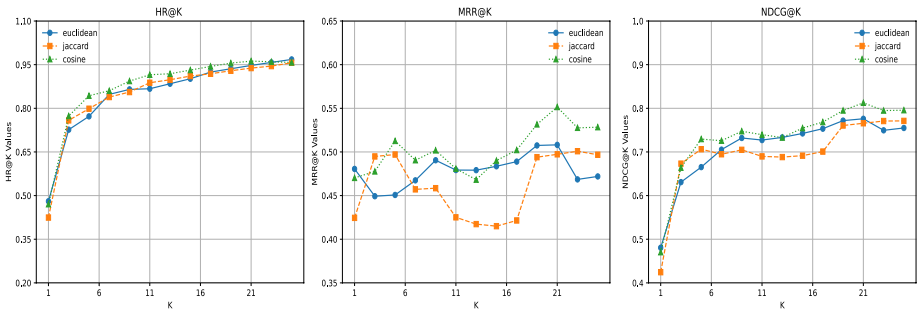

**Fig. 8** Evaluation of similarity functions in the CyberSwarm algorithm: this chart shows how different similarity functions-Euclidean, Jaccard, and Cosine-affect the algorithm's performance. Metrics like HR@k, MRR@k, and NDCG@k assess recommendation accuracy and relevance for each function

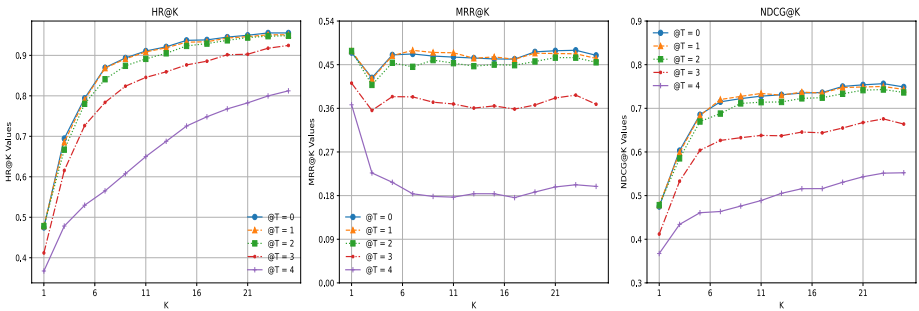

**Fig. 9** Effect of rating threshold filtering on CyS's recommendation performance: this chart shows how different rating thresholds (1 to 5) impact CyberSwarm's performance, highlighting the trade-offs between filtering stringency and recommendation quality and how threshold changes affect accuracy and relevance

### 6.4 Impact of removing isolated nodes on hypergraph integrity and performance

The study highlights the importance of isolated nodes, which, despite lacking connections, enhance recommendation accuracy in the proposed model. Figure 10a shows the impact of removing isolated nodes before input to CyS, evaluated using precision@k, recall@k, and f-score@k. The presence of isolated nodes in the dataset improves recommendation performance, underscoring their contribution to our model's superior recommendation capability.

### 6.5 Effect of centrality measures on recommendation performance

In the feature extraction phase of the CyS, centrality measures-degree, betweenness, and closeness centrality-are essential for improving model performance. Figure 10b demon-strates that incorporating these measures into node embeddings enhances CyS's recommendation accuracy by providing insight into node importance. Centrality quantifies node influence and connectivity, enabling CyS to make more informed recommendations with improved precision by leveraging the structural and functional dynamics within the network.

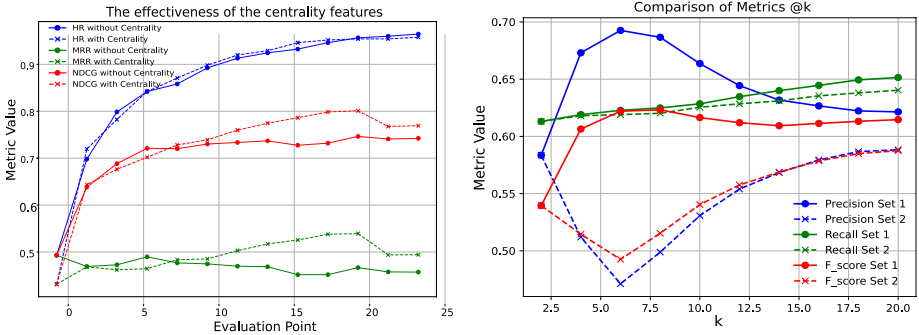

**Fig. 10** Impact of graph modifications on recommendation performance: **a** Comparison of Precision@k, Recall@k, and F-score@k with and without centrality measures in feature representation; **b** Comparison of these metrics before and after removing isolated nodes from the hypergraph

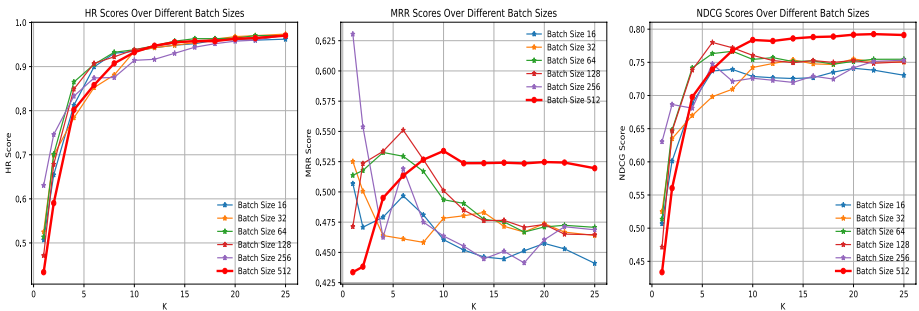

**Fig. 11** Performance comparison of different batch sizes (16, 32, 64, 128, 256, and 512) across three metrics: HR, MRR, and NDCG as a function of $K$. The batch size of 128 (red line) consistently outperforms other batch sizes across all metrics, demonstrating its effectiveness in achieving higher recommendation quality

### 6.6 Effect of batch size on recommendation performance

Figure 11 presents the performance of various batch sizes (16, 32, 64, 128, 256, and 512) on three key metrics: HR, MRR, and NDCG. The chart of the HR (Left subplot) shows that the 512 batch size achieves the highest HR scores, especially as $K$ increases, indicating its effectiveness in maximizing relevant hits. Batch sizes 32 and 64 perform well but remain behind 512. Dealing with MRR (Middle subplot), 512 remains near the top across $K$ values. But in NDCG (Right subplot), Similar to HR, the 512 batch size leads in NDCG performance, with 128 and 64 trailing closely but unable to match 128, especially at high $K$ values. The smallest (16) and largest (256) batch sizes show lower performance, underscoring the limitations of very small or large batch sizes in this setting.

### 6.7 Effect of Node2Vec size on recommendation performance

As shown in Fig. 12a, the HR scores generally increase as $K$ rises, with the red line (embedding size 64) consistently leading and demonstrating superior performance across most values of $K$. This embedding size maintains a higher and more stable HR score compared to others, even as the performance of all sizes tends to plateau around $K = 10$. Figure 12b, which shows MRR scores, reveals more variability with different embedding sizes. The red line (embedding size 64) maintains relatively stable performance but shows some decline at higher $K$ values. Unlike HR, the MRR scores fluctuate more, indicating that different embedding sizes may excel at different $K$ levels. Figure 12c depicts NDCG scores, which generally follow an upward trend similar to HR as $K$ increases. Here again, the embedding size 64 outperforms the other sizes across most $K$ values, while embedding sizes like 16 and 32 remain competitive, particularly at lower $K$ values. These results show the effect of the N2Vec size on the overall results and its contribution to improving performance.

### 6.8 Sensitivity analysis of the CyS performance with respect to $\lambda_1$, $\lambda_2$, and $\lambda_3$

To assess the robustness of our model with respect to the centrality weight parameters $(\lambda_1, \lambda_2, \lambda_3)$, we conducted an extensive sensitivity analysis using a grid search across multiple values for each weight. Specifically, we evaluated all combinations from the set $\{0.1, 0.3, 0.6, 0.9\}^3$, while measuring HR and NDCG over a range of top-$k$ values. The results, summarized in Fig. 13, indicate that the model exhibits stable and consistent behavior across a wide range of weight configurations. This confirms that the system is not overly sensitive to specific parameter choices, thereby reinforcing the reproducibility and generality of our approach. While the model performs robustly across settings, we observed marginal gains when emphasizing Degree centrality, suggesting its relatively stronger contribution in guiding the swarm dynamics under our evaluation scenario.

## 7 Conclusion and future work

The proposed algorithm introduces an advancement in recommendation systems by combining swarm intelligence with the intricate dynamics of cyber-social networks. Drawing inspiration from social psychology, it models the evolution of user preferences and com-munity influences within a dynamic hypergraph framework. Each node represents an individual or entity, with real-time interactions and preferences shaped by centrality metrics and Node2Vec embeddings. This unique combination enables the system to dynamically adapt to changing user behaviors and provide recommendations that reflect both individual interests and broader social influences.

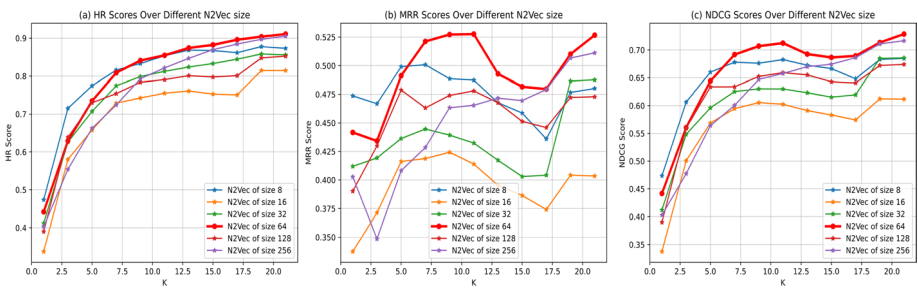

**Fig. 12** Illustrates the performance of various embedding sizes (8, 16, 32, 64, 128, 256) across three key evaluation metrics: HR, MRR, and NDCG. Each subplot represents one of these metrics, with the x-axis showing different values of K and the y-axis depicting the respective metric scores

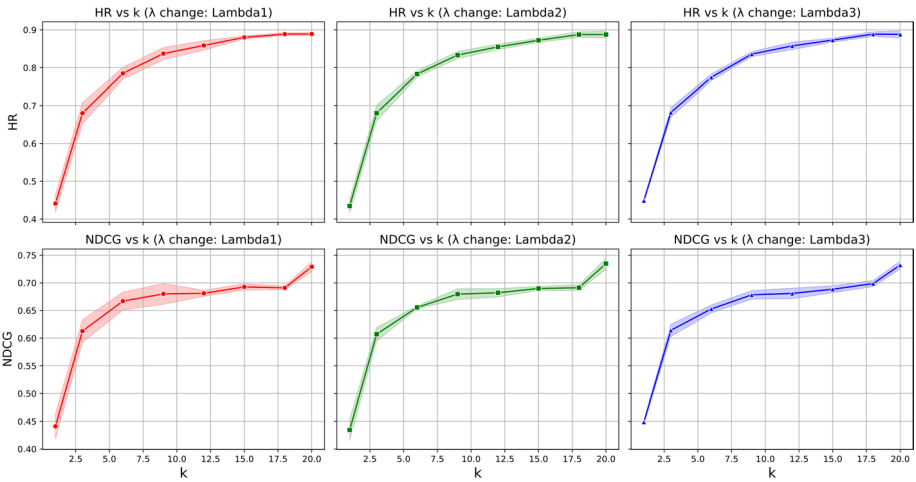

**Fig. 13** Sensitivity analysis of the CyS performance with respect to $\lambda_1$, $\lambda_2$, and $\lambda_3$. Each column varies one parameter while holding the others fixed at 0.1. Performance metrics (HR and NDCG) are reported across different values of $k$

Our proposed CyS algorithm outperforms 38 baseline models across various domains and key metrics, including Hit Rate (HR), Mean Reciprocal Rank (MRR), and Normalized Discounted Cumulative Gain (NDCG). For instance, in the Ciao dataset, it achieves an HR@1 score of 0.4649, a remarkable 38.87% improvement over the second-best method, DANSER. This superiority is evident across all key metrics, with CyS demonstrating a 46.32% improvement in NDCG@20 compared to other models. In the Epinions dataset, it achieves an HR@1 score of 0.4237, marking a 14.04% improvement over the second-best model, HC-CED. The most significant improvement is seen in NDCG@20, where our algorithm scores 0.7117, outpacing the next best model by a substantial 26.16%. Similarly, on the FilmTrust dataset, our model leads across all metrics. Its HR@1 score of 0.5831 is 3.36% higher than that of the second-best model, HC-CED. The most notable improvement is again seen in NDCG@20, where CyS's score of 0.7563 surpasses the closest competitor by 5.6%. In the Gowalla dataset, the proposed model dominates with noticeable higher scores across the key metrics. It achieves an HR@10 score of 0.4888 and an HR@20 score of 0.5827, substantially outperforming the second-best model, HyperS2Rec, which scored 0.4596 and 0.5420, respectively. Furthermore, the our model excels in MRR and NDCG metrics, with particularly strong gains in MRR@10 (0.3912) and NDCG@20 (0.4512), outperforming second-best models FSM and SERec by considerable margins. In the Brightkite dataset, CyS continues its dominance, achieving an HR@10 score of 0.7467 and an NDCG@10 score of 0.7258. These scores are significantly higher than those of the second-best model, SoNeuMF, which scored 0.6588 and 0.4563, respectively. Overall, these results reinforce the proposed algorithm's superior ability to retrieve and rank relevant items across a wide variety of datasets.

As a general-purpose optimization approach, the method is designed to go beyond conventional recommendation tasks. By modeling preference evolution through hierarchical graph structures and incorporating features like degree, closeness, and betweenness centrality, it effectively captures both local interactions and global trends. This flexibility makes it suitable for applications across diverse domains, such as social networking, personalized learning, and medical graph analysis. Its ability to adapt to shifting conditions ensures that recommendations remain relevant and precise over time. The adaptive framework excels in balancing individual preferences with collective social dynamics. By leveraging dynamic feature extraction and message-passing mechanisms, the algorithm uncovers meaningful patterns in sparse or noisy data. This context-aware approach makes it capable of handling complex relationships within a network, ensuring that influential nodes are weighted appropriately while also considering the nuanced behaviors of less connected nodes. Such a balance contributes to its exceptional performance in recommendation tasks. This work establishes a foundation for versatile and adaptive recommendation systems. Its integration of centrality-driven insights, hypergraph modeling, and psychological principles marks a significant step forward in understanding and leveraging network dynamics. By bridging the gap between individual preferences and community influences, the approach offers a powerful tool for advancing personalized recommendations across diverse applications. This innovation not only enhances the accuracy of recommendations but also opens the door to broader applications in other graph-based optimization challenges.

To maximize our algorithm's potential, future work could focus on tailoring it to domain-specific applications, where the algorithm's general-purpose foundation could be enhanced with targeted data and methods. In healthcare, for example, integrating clinical histories could enable it to make precise, personalized treatment recommendations. In e-learning, it could support adaptive learning paths by incorporating learning analytics, helping students receive tailored guidance. Additionally, by conducting focused research into fields like pharmaceuticals for drug discovery or e-commerce for product recommendations, the proposed algorithm could address unique industry needs through specific datasets and fine-tuned configurations.

Our proposed Algorithm also has the potential to extend beyond single-domain recommendations. By handling cross-domain data, it could generate recommendations that draw insights from one area, like a user's movie preferences, to inform suggestions in a related area, like books or courses. Moreover, utilizing multi-modal data, including text, images, and audio, could allow this algorithm to create comprehensive node profiles for more precise recommendations in content-rich environments. Integrating advances in explainable AI would enhance CyS's interpretability, making its recommendations transparent and understandable to nodes. As CyS adapts to large-scale language models and processes complex, unstructured data, it could deliver even more accurate, insightful recommendations across a wide range of fields and applications.

**Funding** Open access funding provided by The Science, Technology & Innovation Funding Authority (STDF) in cooperation with The Egyptian Knowledge Bank (EKB).

**Data availability** The datasets used in this study are publicly available: 1- The Gowalla and Brightkite datasets, containing user check-ins in location-based social networks, were introduced in https://doi.org/10.1145/2020408.2020579. 2- The FilmTrust dataset, which includes movie ratings and trust relationships, was described in https://doi.org/10.1145/2856037. 3- The Epinions dataset, incorporating user ratings, reviews, and trust relationships, was presented in https://doi.org/10.1007/978-3-540-30468-5_31. 4- The Ciao dataset, containing user reviews and social connections, was detailed in https://dl.acm.org/doi/10.1145/2124295.2124309. Readers may refer to these sources for dataset access and further details. 5- The Hetionet dataset, a heterogeneous network integrating biomedical knowledge such as genes, diseases, and drugs, was presented in https://doi.org/10.7554/eLife.26726.

## Declarations

**Conflict of interest** The authors declare no Conflict of interest

**406** Page 48 of 49  A. Elfergany et al.

## Authors and Affiliations

**Abdelsadeq Elfergany[1] · Ammar Adl[1] · Mohammed Kayed[1]**

✉ Ammar Adl
ammar@fcis.bsu.edu.eg; ammaradl@gmail.com

Abdelsadeq Elfergany
a.k.sadeq@fcis.bsu.edu.eg

Mohammed Kayed
kayed@fcis.bsu.edu.eg

[1] Faculty of Computers and Artificial Intelligence, Beni-Suef University, Beni-Suef, Egypt